\documentclass[aps,prd,twocolumn,superscriptaddress,preprintnumbers,floatfix,nofootinbib,notitlepage,showkeys]{revtex4-1}

\usepackage[utf8]{inputenc}
\usepackage{cancel}

\usepackage{graphicx}
\usepackage{hyperref}
\usepackage{latexsym}
\usepackage{amsmath}
\usepackage{amssymb}
\usepackage{bbm}

\usepackage{ulem}
\usepackage{pdfsync}
\usepackage{epsfig}
\usepackage{epstopdf}
\usepackage{subfigure}
\usepackage{color}
\usepackage{comment}
\usepackage{slashed}

\def\beq{\begin{equation}}
\def\eeq{\end{equation}}
\def\baq{\begin{eqnarray}}
\def\eaq{\end{eqnarray}}

\newcommand{\be}{\begin{equation}} % only untightened
\newcommand{\ee}{\end{equation}}
\newcommand{\bea}{\begin{eqnarray}} % only untightened
\newcommand{\eea}{\end{eqnarray}}

\newcommand{\bmp}{\noindent\begin{minipage}{16cm}}
\newcommand{\emp}{\end{minipage}\vskip 7mm} % 7mm untightened
\def\lsim{\mathrel{\raise.3ex\hbox{$<$\kern-.75em\lower1ex\hbox{$\sim$}}}}
\def\gsim{\mathrel{\raise.3ex\hbox{$>$\kern-.75em\lower1ex\hbox{$\sim$}}}}

\newcommand{\intron}[1]{}%{#1}

\begin{document}

\title{Trans-Planckian Censorship, Inflation and Dark Matter}

\author{Tommi Tenkanen}
\email{ttenkan1@jhu.edu}
\affiliation{Department of Physics and Astronomy, Johns Hopkins University, \\
Baltimore, MD 21218, USA}

\begin{abstract}

If the inflationary phase lasted longer than the minimal period, the length scales observed today originate from modes that were smaller than the Planck length during inflation. It was recently argued that this "trans-Planckian problem" can never arise in a consistent string theory framework, which places a stringent constraint on the energy scale of inflation, $V^{1/4}\lesssim 10^9$ GeV. In this paper, we show that this requirement corresponds to a very small Hubble scale during inflation, $H_{\rm inf}\lesssim 1$ GeV, and therefore has serious consequences on scenarios where the dark matter density was generated by amplification of quantum fluctuations during inflation. We also present a class of inflationary models which both satisfy the above limit for the scale of inflation and are in perfect agreement with observational data.
\end{abstract}

%%%%%%%%%%%%%%%%%%%%%%%%%%%%%%%%%%%%%%%%%%%%%%%%%%%%%%%%%%%%%%%%%%%%%%%%%%%%%%%%%%%%%%%%%%%%%%%%%%%%
% DOCUMENT
%
\maketitle
%%%%%%%%%%%%%%%%%%%%%%%%%%%%%%%%%%%%%%%%%%%%%%%%%%%%%%%%%%%%%%%%%%%%%%%%%%%%%%%%%%%%%%%%%%%%%%%%%%%%

\section{Introduction}
\label{introduction}

The current paradigm for explaining the origin of the small inhomogeneities that later became the seeds for the large scale structure of the Universe is cosmic inflation, an era of accelerated expansion before the standard Hot Big Bang state of the Universe~\cite{Starobinsky:1980te, Sato:1980yn, Guth:1980zm, Linde:1981mu, Albrecht:1982wi, Linde:1983gd}. Many of its phenomenological realizations have been studied over the past few decades (for a review, see e.g. Ref. \cite{Martin:2013tda}) and its status has only been strengthened by the recent measurements of the Cosmic Microwave Background (CMB) radiation by the Planck satellite \cite{Akrami:2018odb,Chowdhury:2019otk}.

Likewise, the existence of dark matter (DM) seems indisputable \cite{Bertone:2016nfn,Aghanim:2018eyx}. In particular, observations of the CMB and formation of structures at different scales have revealed that DM must be cold and collisionless (non-interacting) and perturbations in its energy density must overlap with those in baryon and radiation energy densities at cosmological scales to a high accuracy.

Yet the origin of dark matter remains unknown. Instead of undergoing usual thermal freeze-out (see e.g. Ref. \cite{Kolb:1990vq}) or e.g. non-thermal freeze-in \cite{McDonald:2001vt,Hall:2009bx,Bernal:2017kxu} during the Hot Big Bang epoch, the observed dark matter abundance may have been initiated purely gravitationally either during or after cosmic inflation \cite{Ford:1986sy,Turner:1987vd,Kolb:1998ki,Chung:1998zb,Peebles:1999fz,Nurmi:2015ema,Graham:2015rva,Markkanen:2015xuw,Bertolami:2016ywc,Cosme:2017cxk,Alonso-Alvarez:2018tus,Fairbairn:2018bsw,Markkanen:2018gcw,Tenkanen:2019aij}. While sometimes dubbed as a 'nightmare scenario' due to its limited testability, this is a serious possibility which should be studied exhaustively. In fact, some promising ideas about testing very weakly coupled dark matter candidates whose abundance was generated during inflation have recently been presented in the literature \cite{Graham:2015rva,Alonso-Alvarez:2018tus,AlonsoAlvarez:2019cgw,Tenkanen:2019aij}, all of them suggesting to study the properties of DM with observations of formation of structures at different scales (see e.g. Ref. \cite{Amendola:2016saw}). 

However, while inflation is in principle successful in explaining the initial density perturbations at different scales, there exists a theoretical problem. In Refs. \cite{Martin:2000xs,Brandenberger:2000wr,Brandenberger:2012aj} it was realized that if the inflationary phase lasted somewhat longer than the minimal period, then the length scales we observe today originate from modes that were smaller than the Planck length during inflation. This was called the \textit{trans-Planckian problem}, as in that case the usual computations of the perturbation spectrum involve extrapolating low energy physics into regions where it is not necessarily applicable. 

Recently, this lead the authors of Ref. \cite{Bedroya:2019snp} to conjecture that the trans-Planckian problem can never arise in a consistent string theory framework and such inflationary models belong to the "Swampland" \cite{Ooguri:2006in,Obied:2018sgi} (see Ref. \cite{Palti:2019pca} for a review). The authors called this the "Trans-Planckian Censorship Conjecture" (TCC) and studied its immediate implications for inflation in Ref. \cite{Bedroya:2019tba} (see also Ref. \cite{Cai:2019hge}).

In this paper, we do not contemplate on the validity of the TCC but instead point out that the restrictions it places on inflationary models are not particularly problematic. We show that it is relatively easy to construct a class of simple inflationary models which not only exhibit a very low scale of inflation but are also in perfect agreement with all observational data. However, we show that the low scale of inflation the TCC anticipates is generically catastrophic for models in which the DM density was generated by amplification of quantum fluctuations during inflation. This conclusion is independent of the validity of the TCC and only relies on assumptions of the low scale and small duration of inflation, as we will show, and therefore interesting also in its own right. In regards to the TCC, howeer, we point out few novel aspects that were not considered in Refs. \cite{Bedroya:2019snp,Bedroya:2019tba}.

The paper is organized as follows: in Sec. \ref{conjecture}, we revisit the arguments given in Ref. \cite{Bedroya:2019tba}, discuss what the TCC requires from inflationary models, and also provide some novel results. In Sec. \ref{inflation}, we present a class of simple models that satisfy the TCC by exhibiting a low scale of inflation. In Sec. \ref{darkmatter}, we discuss the implications of low-scale inflation for very weakly coupled scalar dark matter models. In Section \ref{conclusions}, we conclude.

%%%%%%%%%%%%%%%%%%%%%%%%%%%%%%%%%%%%%%%%%%%%%%%%%%%%%%%%%%%%%%%%%%%%%%%%%%%%%%%%%%%%%%%%%%%%%%%

\section{The Trans-Planckian Problem}
\label{conjecture}

If the inflationary phase lasted somewhat longer than the minimal period, then the length scales we observe today originate from modes that were smaller than the Planck length during inflation. According to the TCC, this is not allowed, because no length scales which exited the Hubble horizon during inflation could ever have had a wavelength smaller than the Planck length. This amounts to requiring
\begin{equation}
\label{TCC_cond}
\frac{a_{\rm end}}{a_{\rm ini}}\frac{1}{M_{\rm P}} < \frac{1}{H_{\rm end}} ,
\end{equation}
where $a_{\rm ini}$ ($a_{\rm end}$) is the scale factor at the beginning (end) of inflation, $M_{\rm P}$ is the reduced Planck mass, and $H_{\rm end}$ is the Hubble parameter at the end of inflation. In the following, we assume $H$ remains constant during inflation, $H_{\rm inf}=H_{\rm end}$, where $H_{\rm inf}$ denotes the overall scale of inflation (and, hence, $1/H_{\rm inf}$ is the horizon during inflation). This turns out to be an excellent approximation for models we will consider, as will be discussed below. 

Defining then the usual number of e-folds $N\equiv \rm{ln}(a_{\rm end}/a_{\rm ini})$ allows us to express the condition \eqref{TCC_cond} as
\begin{equation}
\label{N_cond}
e^N < \frac{M_{\rm P}}{H_{\rm end}} = \frac{\sqrt{3}M_{\rm P}^2}{\sqrt{V}} ,
\end{equation}
where we used the Friedmann equation in the slow-roll approximation, $\rho_{\rm tot}\simeq V = 3H_{\rm inf}^2M_{\rm P}^2$, where $\rho_{\rm tot}$ is the total energy density governed by $V$, the inflaton potential energy during inflation. 

As Eq. \eqref{N_cond} shows, the TCC constrains the duration of inflation. Indeed, if the inflationary phase lasted for long enough, then the length scales we observe today originated from modes that were smaller than the Planck length during inflation. On the other hand, we want inflation to last for long enough to explain e.g. the well-known horizon problem (see e.g. Ref. \cite{Baumann:2018muz}). That is, we require
\begin{equation}
\label{N_minimum}
\frac{1}{H_{\rm inf}}\,e^N\, \frac{a_{\rm reh}}{a_{\rm end}}\frac{a_0}{a_{\rm reh}} \geq \frac{1}{H_0} ,
\end{equation}
where $a_{\rm reh}$ denotes the time of reheating after inflation and $a_0$ is the scale factor today. Due to entropy conservation, we have 
\begin{equation}
\frac{a_0}{a_{\rm reh}} = \left(\frac{g_{*S}(T_{\rm reh})}{g_{*S}(T_0)}\right)^{1/3}\frac{T_{\rm reh}}{T_0}\,,
\end{equation}
where $T_{\rm reh}$ is the reheating temperature, $T_0$ is the present CMB temperature and $g_{*S}$ denote the effective number of entropy degrees of freedom. By assuming instant reheating, we can use $a_{\rm reh} = a_{\rm end}$ and
\begin{equation}
T_{\rm reh} = \left(\frac{30}{\pi^2 g_*(T_{\rm reh})}\right)^{1/4} V^{1/4} ,
\end{equation}
so that together with Eq. \eqref{N_minimum}, the condition \eqref{N_cond} gives an upper limit on the energy scale of inflation
\begin{equation}
\frac{V^{1/4}}{M_{\rm P}} < \left(\frac{g_{*S}(T_{\rm reh})}{g_{*S}(T_0)}\right)^{1/9}\left(\frac{30}{\pi^2g_*(T_{\rm reh})}\right)^{1/12}\left(\frac{3H_0}{T_0}\right)^{1/3}\,. 
\end{equation}
Using then the observed CMB temperature $T_0 = 2.725$ K and $H_0 = h\times 2.13\times 10^{-42}$ GeV, we obtain
\begin{equation}
\label{TCC_limit}
V^{1/4} < 8\times 10^8 h^{1/3} {\rm GeV} \equiv V_{\rm max}\,,
\end{equation}
where $V_{\rm max}$ is defined for later purposes. Taking $h=0.68$ \cite{Aghanim:2018eyx} gives $V^{1/4} < 7\times 10^8$ GeV. This result agrees with the original calculation in Ref. \cite{Bedroya:2019tba} when all factors of the order unity are included.

It is instructive to see what the upper limit on the energy scale of inflation \eqref{TCC_limit} implies for observables. In particular, the tensor-to-scalar ratio is defined as
\begin{equation}
\label{r_def}
r \equiv \frac{\mathcal{P}_T(k_*)}{\mathcal{P}_\zeta(k_*)} = \frac{8}{M_{\rm P}^2}\left(\frac{H_{\rm inf}}{2\pi}\right)^2 \mathcal{P}_\zeta(k_*) ,
\end{equation}
where $\mathcal{P}_T, \mathcal{P}_\zeta$ are the tensor and scalar curvature power spectra, respectively, measured at the pivot scale $k_*$. In the following, we take $k_*=0.002\, {\rm Mpc}^{-1}$. By exchanging $H_{\rm inf}$ with $r$ in the Friedmann equation, the energy scale can then be expressed as
\begin{equation}
\frac{V}{M_{\rm P}^4} = \frac{3\pi^2 \mathcal{P}_\zeta(k_*)r}{2} ,
\end{equation}
so that
\begin{equation}
\label{rmax}
r = \frac{2}{3\pi^2}\mathcal{P}_\zeta^{-1}(k_*)\frac{V}{M_{\rm P}^4} < 2\times 10^{-31} \equiv r_{\rm max},
\end{equation}
where we used $\mathcal{P}_\zeta(k_*) = 2.1\times 10^{-9}$ and $h=0.68$ \cite{Aghanim:2018eyx}, and $r_{\rm max}$ is again defined for later purposes. 

The limit \eqref{rmax} is considerably lower than what is expected to be detected or ruled out by the next generation CMB B-mode polarization experiments such as BICEP3 \cite{Wu:2016hul}, LiteBIRD~\cite{Matsumura:2013aja} and the Simons Observatory \cite{Simons_Observatory}, which aim at $r\sim 0.001-0.01$. Therefore, as concluded in Ref. \cite{Bedroya:2019tba}, the result \eqref{rmax} shows that any detection of primordial tensor perturbations on cosmological scales would provide evidence for a different origin of the primordial tensor perturbation spectrum than any inflationary model consistent with the TCC. 

Before concluding this section, we make some remarks which were not discussed in Refs. \cite{Bedroya:2019snp,Bedroya:2019tba}. First, the results \eqref{r_def} and \eqref{rmax} also give
\begin{equation}
\label{H_limit}
\frac{H_{\rm inf}}{\rm GeV} \leq 0.25\sqrt{\frac{r}{10^{-30}}} ,
\end{equation}
which is particularly interesting for models where dark matter originated from inflation or for curvaton-like models \cite{Enqvist:2001zp,Lyth:2001nq,Moroi:2001ct}, because $H_{\rm inf}$ controls the magnitude of scalar field fluctuations during inflation, were these scalar fields driving inflation or not. This will be discussed in more detail in Sec. \ref{darkmatter}. 

The upper limit \eqref{TCC_limit} has also implications for the total duration of inflation. From Eq. \eqref{N_cond}, we obtain
\begin{equation}
\label{N_total}
N < {\rm ln}\left(\frac{\sqrt{3}M_{\rm P}^2}{\sqrt{V}}\right)\,.
\end{equation}
Applying then the constraint \eqref{TCC_limit} on $V$, we see that the maximum number of e-folds allowed by the largest $V$ is 
\begin{equation}
\label{Nlimit_Vmax}
N(V_{\rm max}) \simeq 45\,,
\end{equation}
where $V_{\rm max}$ is given by Eq. \eqref{TCC_limit}. This is of course consistent with the value that solves the horizon problem, as the derivation of $V_{\rm max}$ required it, see Eq. \eqref{N_minimum}. However, such a low value of $N$ has implications for inflationary models, as we shall see in the next section. 

There is also a lower limit on $V$, which comes from Big Bang Nucleosynthesis (BBN). Requiring 
\begin{equation}
V > \frac{\pi^2}{30}g_{*}(T_{\rm BBN})T_{\rm BBN}^4\,,
\end{equation}
where $T_{\rm BBN}\simeq 4$ MeV \cite{Kawasaki:2000en,Hannestad:2004px,Ichikawa:2005vw,DeBernardis:2008zz} and $g_{*}(T_{\rm BBN})=10.75$, we obtain $V^{1/4}_{\rm min} \simeq 5$ MeV. This gives
\begin{equation}
\label{Nlimit_Vmin}
N(V_{\rm min}) \simeq 96\,,
\end{equation}
which, according to the TCC, is the strict upper limit on the total duration of inflation. While this limit is far larger than the usual number $N \sim 60$ required to solve the horizon problem\footnote{For $V^{1/4}_{\rm min}$, the required value of $N$ that solves the horizon problem is of course much smaller, $N\sim 22$.} and may therefore seem  to be mainly of theoretical interest, it has serious consequences for certain types of dark matter models, as will be discussed in Sec. \ref{darkmatter}.

In the next section, we present a class of models which satisfy the constraints \eqref{TCC_limit} and \eqref{H_limit} for suitable choices of model parameters. Then, in Sec. \ref{darkmatter}, we discuss implications of the TCC for models where the DM density was generated during inflation.

%%%%%%%%%%%%%%%%%%%%%%%%%%%%%%%%%%%%%%%%%%%%%%%%%%%%%%%%%%%%%%%%%%%%%%%%%%%%%%%%%%%%%%%%%%%%%%%

\section{Low-scale inflation}
\label{inflation}

The class of models we will consider were recently introduced in Refs. \cite{Enckell:2018hmo,Antoniadis:2018ywb} (see also Refs. \cite{Antoniadis:2018yfq,Tenkanen:2019jiq}). The general action for the models reads
\begin{equation}
\label{SJ}
S_J = \int {\rm d}^4x \sqrt{-g}\left(\frac12 F(R,\phi)- \frac12g^{\mu\nu}\partial_\mu\phi\partial_\nu\phi - V(\phi)\right) ,
\end{equation}
where $g$ is the determinant of the metric $g_{\mu\nu}$, $R=g^{\mu\nu}R^\lambda_{\mu\lambda\nu}(\Gamma,\partial\Gamma)$ is the curvature (Ricci) scalar, $R^\lambda_{\mu\lambda\nu}$ is the Riemann tensor constructed from the connection $\Gamma$ and its derivatives, and $\phi$ is the inflaton field. In the following, we take
\begin{equation}
\label{F}
F(R,\phi) = M_{\rm P}^2 R + \alpha R^2 +G(\phi)R,
\end{equation}
where $\alpha$ is a dimensionless parameter and $G(\phi)$ encapsulates the possible non-minimal couplings between the inflaton and gravity. 

The action \eqref{SJ} is particularly generic, as both the non-minimal coupling and the Starobinsky-like $\alpha R^2$ term can be argued to be generated by quantum corrections in a curved background \cite{Birrell:1982ix}. However, here we make a slightly unusual choice by assuming \textit{Palatini} gravity, where both the spacetime metric $g_{\mu\nu}$ and the connection $\Gamma$ are treated as independent variables; see Ref. \cite{Tenkanen:2020dge} for an introduction to the topic. As was shown in Refs. \cite{Enckell:2018hmo,Antoniadis:2018ywb,Antoniadis:2018yfq,Tenkanen:2019jiq} and as will be discussed below, this has interesting consequences, as the curvature scalar $R$ is a function of both $g_{\mu\nu}$ and $\Gamma$, whereas the kinetic term for the inflaton only depends on the metric $g_{\mu\nu}$. As a result, the energy scale of inflation can be made arbitrarily low.

Note that one can take this approach also in the context of the General Relativity, where no non-minimal couplings exist and the $\alpha R^2$ term is absent. In that case the constraint equation for the connection renders the two theories equivalent, i.e. in that case the metric theory (i.e. the one where there is a unique $\Gamma=\Gamma(g_{\mu\nu})$, the \textit{Levi-Civita} connection) and Palatini theories are nothing but two different formulations of the same theory. However, with non-minimally coupled matter fields (i.e. with $G(\phi)\neq 0$) or otherwise enlarged gravity sector (i.e. with $\alpha \neq 0$) this is generally not the case~\cite{Sotiriou:2008rp}. Therefore, in the context of non-minimal models, such as the one in Eq. \eqref{F}, one has to make a \textit{choice} of the underlying gravitational degrees of freedom. 

Note that choosing the Palatini approach does not constitute a modified theory of gravity any more than the metric one does, as currently we do not know what the fundamental gravitational degrees of freedom are. Also note that this choice does not necessarily amount to adding new degrees of freedom to the theory. In fact, with our choice of $F(R,\phi)$, Eq. \eqref{F}, there are \textit{less} dynamical degrees of freedom in the Palatini case than there would be in the metric one \cite{Enckell:2018hmo}. In the Palatini case the action \eqref{SJ} describes single-field inflation, whereas in the metric case it becomes a genuine two-field model (see e.g. Refs. \cite{Enckell:2018uic,Canko:2019mud} for recent work and references therein). The Palatini approach is therefore not only quite general (as it allows $\Gamma$ to be, a priori, independent of $g_{\mu\nu}$) but also very simple. In the following, however, we will make one extra assumption for simplicity: we assume that the connection is torsion-less, $\Gamma^\lambda_{\mu\nu}=\Gamma^\lambda_{\nu\mu}$. For non-vanishing torsion, see Ref. \cite{Rasanen:2018ihz}.

A Weyl transformation
\begin{equation}
g_{\mu\nu}\to \left(\varphi + G(\phi)\right)g_{\mu\nu} ,
\end{equation}
where $\varphi$ is an auxiliary field \cite{Enckell:2018hmo} then shows that the action \eqref{SJ} is equivalent to 
\begin{eqnarray}
\label{SE}
S_E &=& \int {\rm d}^4x \sqrt{-g}\bigg[\frac12 M_{\rm P}^2 R- \frac12\partial^\mu\chi\partial_\mu\chi \\ \nonumber 
&+& \frac{\alpha}{2}\left(1+8\alpha\frac{\bar{U}}{M_{\rm P}^4}\right)\left(\partial^\mu\chi\partial_\mu\chi\right)^2 - U(\chi)\bigg] ,
\end{eqnarray}
where the re-defined inflaton field $\chi$ is given by
\begin{equation}
\frac{{\rm d}\phi}{{\rm d}\chi} =\sqrt{\left(1+G(\phi)\right)\left(1+8\alpha\frac{\bar{U}}{M_{\rm P}^4}\right)} ,
\end{equation}
and the potential is
\begin{eqnarray}
U(\chi) &\equiv& \frac{\bar{U}(\chi)}{1+8\alpha\bar{U}(\chi)/M_{\rm P}^4}\,, \\ \nonumber 
\bar{U}(\chi) &\equiv& \frac{V(\phi(\chi))}{\left(1+G(\phi(\chi)) \right)^2} .
\end{eqnarray}
Because the action \eqref{SE} has a canonical gravity sector, we say that the theory is defined in the Einstein frame, in contrast to the Jordan frame \eqref{SJ} where the gravity sector is non-canonical. We see that in addition to modifying the potential, the Weyl transformation has translated the $R^2$ term into a higher order kinetic term for the inflaton field. Because a negative $\alpha$ would lead to negative kinetic energy, we assume $\alpha \geq 0$.

Given the Einstein frame action \eqref{SE} and the canonically normalized\footnote{It was shown in Ref. \cite{Enckell:2018hmo} that the $(\partial\chi)^4$ term in Eq. \eqref{SE} is negligible in slow-roll.} scalar field $\chi$, one can then derive the equations of motion in the usual way. Slow-roll inflation is described by the usual slow-roll parameters
\begin{eqnarray}
\epsilon &\equiv& \frac12 M_{\rm P}^2\left(\frac{U_{,\chi}}{U}\right)^2 =\frac{1}{2} \left(\frac{U_{,\phi}}{U} \frac{{\rm d}\phi}{{\rm d}\chi} \right)^2 = \frac{U}{\bar{U}}\bar{\epsilon}\,,\\ \nonumber
\eta &\equiv& M_{\rm P}^2\frac{U_{,\chi,\chi}}{U} = \bar{\eta} - 24\alpha U \bar{\epsilon}\,,
\end{eqnarray}
where the subscripts of $U$ denote its derivatives with respect to $\chi$ and $\phi$ as specified after the comma and the overbars denote the slow-roll parameters in the case $\alpha = 0$, when $U=\bar{U}$ and ${\rm d}\phi/{\rm d}\chi = \sqrt{1+G(\phi)}$. The amplitude of the curvature power spectrum and the spectral tilt are then given by
\begin{eqnarray}
24\pi^2M_{\rm P}^4\mathcal{P}_\zeta(k_*) &=& \frac{U}{\epsilon} =\frac{\bar{U}}{\bar{\epsilon}} \\ \nonumber
n_s - 1&=& 2\eta - 6\epsilon = 2\bar{\eta} - 6\bar{\epsilon}\,.
\end{eqnarray}
Therefore, even though the $\alpha R^2$ term modifies the slow-roll parameters, the effect cancels out in $\mathcal{P}_\zeta$ and $n_s$. As shown in Ref. \cite{Enckell:2018hmo}, this is also true for the observables defined as higher order derivatives of the scalar curvature power spectrum, such as the running of the spectral tilt, because the curvature power spectrum remains the
same. However, that is not the case for the tensor power spectrum
\begin{equation}
\mathcal{P}_T = \frac{2}{3\pi^2}\frac{U}{M_{\rm P}^4} = \frac{2}{3\pi^2}\frac{\bar{U}/M_{\rm P}^4}{1+8\alpha\bar{U}/M_{\rm P}^4} \,,
\end{equation}
and, as a result, the tensor-to-scalar ratio becomes
\begin{equation}
\label{r_general}
r = 16\epsilon = \frac{\bar{r}}{1 + 8\alpha \bar{U}/M_{\rm P}^4} \,,
\end{equation}
where $\bar{r}\equiv 16\bar{\epsilon}$. This has interesting consequences; in particular, it allows to construct classes of simple models which have a very low scale of inflation, as $V^{1/4}\propto \sqrt{H_{\rm inf}} \propto r^{1/4}$, and which are therefore compatible with the TCC. It also validates our treatment of constant $H_{\rm inf}$ in Sec. \ref{conjecture}, because $-\dot{H}_{\rm inf}/H_{\rm inf}^2 = \epsilon = r/16 \ll 1$ for large $\alpha$.

\begin{figure}
\begin{center}
\includegraphics[width=.495\textwidth]{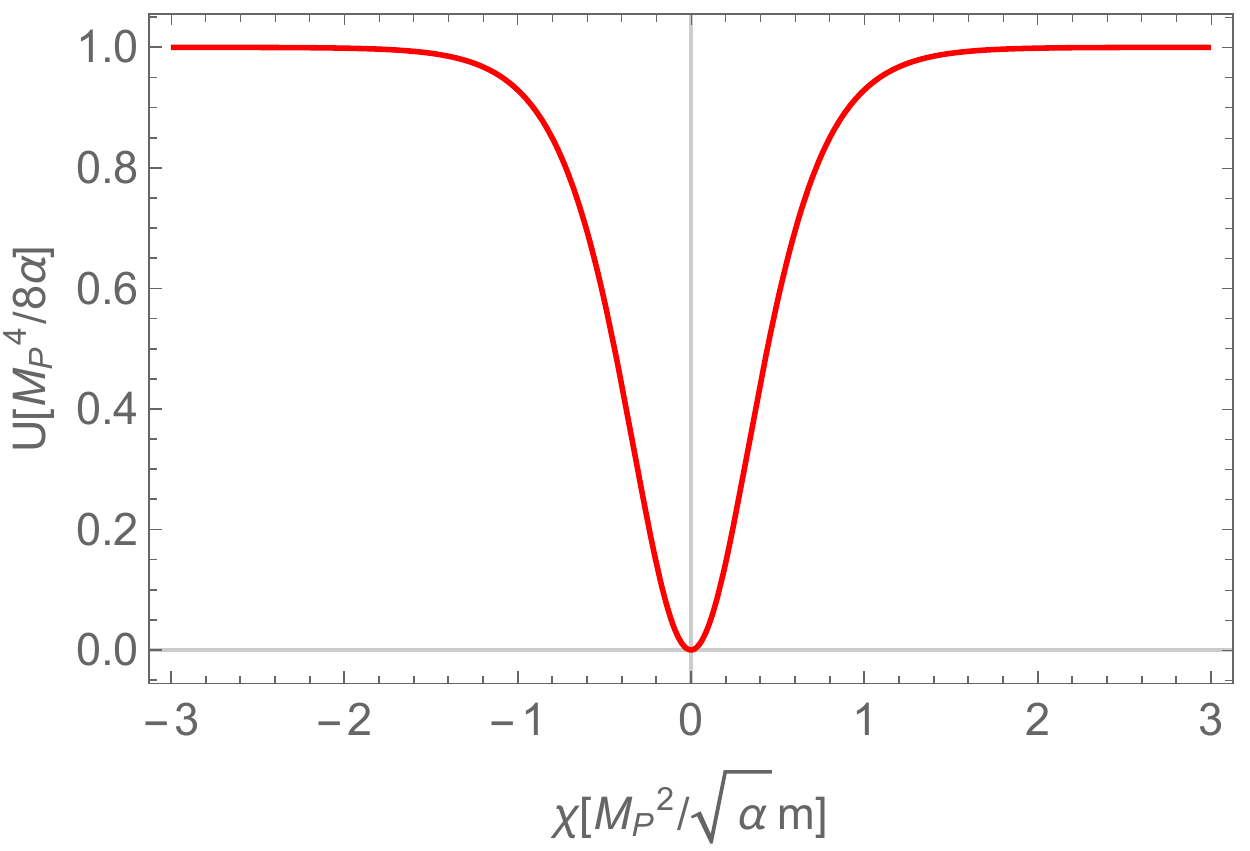}
\caption{The Einstein frame potential for the canonically normalized inflaton field. The potential develops a plateau at $\chi \gtrsim M_{\rm P}^2/\sqrt{\alpha}m$, which is where inflation occurs.}
\label{potential}
\end{center}
\end{figure}

For simplicity, let us consider a free minimally coupled scalar field
\begin{equation}
V(\phi) = \frac12 m^2\phi^2\,, \hspace{.5cm} G(\phi) = 0 ,
\end{equation}
to see what the TCC requires from the model parameters so that the predictions of the model agree with both the TCC and observations of the CMB. Given the above choice of $V(\phi)$ and $G(\phi)$, the Einstein frame potential for the canonically normalized inflaton field $\chi$ becomes
\begin{equation}
U(\chi) = \frac{M_{\rm P}^4}{8\alpha}{\rm tanh}^2\left(2\sqrt{\alpha}\frac{m\chi}{M_{\rm P}^2} \right)\,,
\end{equation}
where $\rm tanh^2(x)\equiv (tanh(x))^2$. The potential is shown in Fig. \ref{potential}. 

The predictions for observables can then be computed in the usual way, as explained above. First, $\bar{n}_s\simeq 1-2/N$ and $\bar{r} \simeq 8/N$ as usual, and
\begin{equation}
\mathcal{P}_\zeta = \frac{32}{3\pi^2\bar{r}^2}\left(\frac{m}{M_{\rm P}}\right)^2 .
\end{equation}
Then, for large $\alpha$ the result \eqref{r_general} becomes 
\begin{equation}
r \simeq \frac{1}{12\pi^2\mathcal{P}_\zeta \alpha} \,,
\end{equation}
whereas 
\begin{equation}
n_s = \bar{n}_s \simeq 1-\frac{2}{N} \,,
\end{equation}
which gives\footnote{The fact that this value is only marginally allowed by the CMB observations is not a huge problem, as in the presence of even a small non-zero gravity coupling $G(\phi)\propto\xi \phi/M_{\rm P}$ with $\xi\gtrsim 0.1$, the prediction for the spectral tilt becomes shifted by a small factor, $n_s \simeq 1-3/(2N)$ \cite{Jarv:2017azx}, which gives $n_s\simeq 0.967$ for $N=45$. This is well within the region preferred by Planck.} $n_s \simeq 0.956$ for the maximum number of e-folds considered in Sec. \ref{conjecture}, $N=N(V_{\rm max})\sim 45$. Requiring $r <  r_{\rm max}$ gives
\begin{equation}
\label{alpha_limit}
\alpha > \frac{1}{12\pi^2\mathcal{P}_\zeta r_{\rm max}} \simeq 2\times 10^{37} \,,
\end{equation}
which is the limiting value of $\alpha$ that makes the scenario compatible with the TCC by lowering the scale of inflation below the limit \eqref{TCC_limit}.

The requirement that one of the key parameters of the model has to take a value this large may at first seem problematic, as one may be concerned about fine-tuning. However, what we effectively have done is introducing a new scale\footnote{This is because had we started in the Einstein frame and simply shifted $U \to \bar{U}/(1+\bar{U}/M^4)$ and introduced a non-canonical extension $\frac{\alpha}{2}(1+\bar{U}/M^4)(\partial^\mu\chi\partial_\mu\chi)^2$, the result would have been the same. This reflects the fact that Palatini theories are \textit{metric-affine} \cite{Azri:2017uor,Azri:2018gsz,Rasanen:2018ihz,Shimada:2018lnm}.} $M \equiv M_{\rm P}/\alpha^{1/4}$, which in the case of $r=r_{\rm max}$ is $M\simeq 10^{-10}M_{\rm P}\simeq 7\times 10^{8}$ GeV, which is roughly the same as the maximum scale of inflation according to the TCC, Eq. \eqref{TCC_limit}. This is not a coincidence, as the flatness of the potential requires $U=\bar{U}/(1+\bar{U}/M^4)\sim M^4 \sim V$, where the last equality comes from the TCC. Whether the existence of this new scale is plausible or not can be argued to depend on the (unknown) UV completion of the theory. Note, however, that allowing for $G(\phi)\neq 0$, for example the usual non-minimal coupling quadratic in the field, $G(\phi) \propto \xi\phi^2$, alleviates the requirement for $\alpha$ for large enough value of the coupling $\xi$. 

Another concern may arise from initial conditions for inflation. It is well known that so-called hilltop inflation models do not exhibit an attractor for trajectories in field space for small field values (see e.g. Ref. \cite{GOLDWIRTH1992223}), and therefore Ref. \cite{Bedroya:2019tba} concluded that if the field range for slow-roll inflation is constrained by the TCC, inflation is faced with an initial condition problem. However, despite the fact that in the model presented above $\chi \ll M_{\rm P}$ for large $\alpha$, this model is not a hilltop or "small-field" model in the usual sense but asymptotes to a plateau model. This class of models is known to {\it not} suffer from the initial conditions problem, as recently discussed in Ref. \cite{Chowdhury:2019otk}. In the model considered in this work, the presence of the $(\partial\chi)^4$ term in Eq. \eqref{SE} complicates the analysis in the more general case -- however, it was recently shown in Ref. \cite{Tenkanen:2020cvw} that plateau models with an $R^2$ term in Palatini gravity indeed do not suffer from the initial conditions problem in the general case. However, as the TCC limits the {\it total} number of inflationary e-folds, it is not enough for the initial conditions to be such that slow-roll inflation is eventually reached. Instead, the conjecture requires slow-roll inflation to begin soon after the initial conditions are set, and therefore provides extra limitations not only to the shape of the inflaton potential but also to the set of possible initial conditions by requiring a small initial velocity $\dot{\chi}$.

Finally, let us discuss some other ways to lower the scale of inflation. Examples include the famous $\alpha$- or $\xi$-attractor models\footnote{See also Ref. \cite{Jarv:2017azx} for the Palatini counterpart of $\xi$-attractors and Refs. \cite{Kaiser:2013sna,Carrilho:2018ffi} for related multifield scenarios.} \cite{Galante:2014ifa}, although especially in the latter case obtaining the correct amplitude for the curvature power spectrum typically requires parameter values which do not make the tensor-to-scalar ratio smaller than $r\sim 10^{-13}$ \cite{Bauer:2008zj,Rasanen:2017ivk,Takahashi:2018brt}. In this sense, inclusion of the $R^2$ term (or more complicated $F(R)$) seems crucial in this context. In any case, the result \eqref{alpha_limit} shows as a proof of principle that it is possible to construct relatively simple inflationary models which satisfy the TCC, and moreover models which predict a tensor-to-scalar ratio far smaller than the canonical value $r\sim \mathcal{O}(0.01-0.1)$.

%%%%%%%%%%%%%%%%%%%%%%%%%%%%%%%%%%%%%%%%%%%%%%%%%%%%%%%%%%%%%%%%%%%%%%%%%%%%%%%%%%%%%%%%%%

\section{Dark matter}
\label{darkmatter}

In this section, we show that the low scale of inflation the TCC anticipates is generically catastrophic for models in which the DM density was generated by amplification of quantum fluctuations during inflation. However, our conclusions will be much more general in a sense that they are independent of the validity of the TCC and only rely on the assumptions of the low scale and small duration of inflation, as will become evident below.

To show this, consider now another scalar field, a \textit{spectator} field, by which we refer to a field which was energetically subdominant during inflation and which did not take part in driving it. For simplicity, we assume the field has a Lagrangian
\begin{equation}
\label{L_DM}
\mathcal{L}_\sigma = -\frac12g^{\mu\nu}\partial_\mu\sigma\partial_\nu\sigma - \frac12 m_\sigma^2\sigma^2 - \frac12\xi_\sigma R \sigma^2.
\end{equation}
where $m_\sigma$ is the bare mass of the field and $\xi_\sigma$ its non-minimal coupling to gravity. In an FRW universe,
\begin{equation}
R = 6\left(\frac{\ddot{a}}{a} + H^2\right)\,,
\end{equation}
so that during inflation $R=12H_{\rm inf}^2$, whereas during radiation-domination when $H=1/(2t)$, the Ricci scalar vanishes, $R=0$. Here we do not attempt to connect this model to the scenario considered in Sec. \ref{inflation} but simply study if the Lagrangian \eqref{L_DM} can give enough DM compatible with the CMB and other observables for a low value of $H_{\rm inf}$. A similar scenario has previously been studied in Refs. \cite{Cosme:2017cxk,Cosme:2018nly,Alonso-Alvarez:2018tus,AlonsoAlvarez:2019cgw} in case of a large inflationary scale. A scenario where the spectator field couples to the inflaton field has been studied in Ref. \cite{Bertolami:2016ywc}, and scenarios where the field exhibits large self-interactions have been studied in Refs. \cite{Peebles:1999fz,Enqvist:2014zqa,Nurmi:2015ema,Kainulainen:2016vzv,Heikinheimo:2016yds,Enqvist:2017kzh,Markkanen:2018gcw}. A free scalar case was studied in Ref. \cite{Tenkanen:2019aij} and a similar scenario in the context of the QCD axion in Refs. \cite{Guth:2018hsa, Graham:2018jyp}.

Let us first assume that at the beginning of inflation, the field was located at the minimum of its potential throughout the inflating domain. However, if the field was light during inflation, $m_{\rm eff}^2 < 9/4 H_{\rm inf}^2$ where $m_{\rm eff}^2 = m_\sigma^2 + \xi_\sigma R$, it still acquired fluctuations during inflation. By splitting the field into $\sigma(\textbf{x},t)=\sigma_0(t) + \delta\sigma(\textbf{x},t)$, expressing the space-dependent part as
\begin{equation}
\delta\sigma(\textbf{x},t) = \int\frac{{\rm d}^3\textbf{k}}{(2\pi)^{3/2}}\left(\delta\sigma_k(t) a_k e^{i\textbf{k}\cdot \textbf{x}}+ \delta\sigma_k^*(t) a_k^\dagger e^{-i\textbf{k}\cdot \textbf{x}}\right)\,,
\end{equation}
where $k=|\textbf{k}|$ and $a_\textbf{k}$ and $a^\dagger_\textbf{k}$ are the usual annihilation and creation operators, respectively, and solving the equation of motion for the mode functions
\begin{equation}
\label{mode_eom}
\delta\ddot{\sigma}_k(t) + 3H_{\rm inf}\delta\dot{\sigma}_k(t) + \left[m_{\rm eff}^2 +\left(\frac{k}{a}\right)^2\right]\delta\sigma_k(t) = 0\,,
\end{equation}
one can show that the fluctuations start accumulating. Assuming the Bunch-Davies vacuum, the fluctuations assume a Gaussian distribution with a variance given by \cite{Bunch:1978yq} (see also Refs. \cite{Vilenkin:1982wt,Turner:1987vd,Starobinsky:1994bd})
\begin{eqnarray}
\label{sigma_fluctuations}
\langle\delta\sigma^2 \rangle &=& \int_{a(t_{\rm ini})H_{\rm inf}}^{a(t_{\rm end})H_{\rm inf}}\frac{{\rm d}^3k}{(2\pi)^{3/2}} |\delta\sigma_k |^2 \\ \nonumber
&=& \frac{3}{8\pi^2}\frac{H^4_{\rm inf}}{m_{\rm eff}^2} \left(1 -e^{-\frac23\frac{m_{\rm eff}^2}{H_{\rm inf}^2}N} \right)\,.
\end{eqnarray}
Therefore, while the classical mean field vanishes, the fluctuations contain energy density and can constitute all or part of dark matter, as recently shown in \cite{Tenkanen:2019aij}. The calculation is slightly more complicated in the self-interacting case but can be performed by solving the Fokker-Planck equation for the distribution of scalar field fluctuations. For details, see Refs. \cite{Starobinsky:1994bd,Markkanen:2019kpv} for the general case and Ref. \cite{Markkanen:2018gcw} in the context of DM.

The result \eqref{sigma_fluctuations} shows that the fluctuations can accumulate only for
\begin{equation}
N \gg \frac32 \frac{H_{\rm inf}^2}{m_{\rm eff}^2}\,.
\end{equation}
If the total duration of inflation is limited, this calls for a large $m_{\rm eff}^2/H_{\rm inf}^2$ ratio. Moreover, essentially the same quantity also controls if the classical background field indeed ends up close to $\sigma_0=0$ or not. By solving the equation of motion for the background field 
\begin{equation}
\ddot{\sigma}_0(t) + 3H_{\rm inf}\dot{\sigma}_0(t) + m^2\sigma_0(t) = 0\,,
\end{equation}
one can see that the background field value approaches zero exponentially fast
\begin{equation}
\sigma_0(t) \simeq e^{-\frac13\frac{m^2_\sigma}{H^2_{\rm inf}}N}\sigma_0(t_{\rm ini}) ,
\end{equation}
where $N = H_{\rm inf}t$ as usual. This would suggest that regardless of its initial value, the background field relaxes to the minimum of its potential in a time scale $N_{\rm rel}\simeq 3H^2_{\rm inf}/m_{\sigma}^2$. The time scale in which the quantum fluctuations start accumulating is exactly half of this, as can be seen from Eq. \eqref{sigma_fluctuations}. However, if we assume the TCC holds, then the maximum limit on the total duration of inflation, Eqs. \eqref{Nlimit_Vmax} and \eqref{Nlimit_Vmin}, strongly constrains the relaxation window\footnote{This is also true for the self-interacting case with $V_\sigma=\frac{\lambda}{4} \sigma^4$, for which a relaxation time scale $N_{\rm rel}\sim 10/\sqrt{\lambda}$ was found in Ref. \cite{Enqvist:2012xn}.}. Indeed, even for $N(V_{\rm min})\simeq 96$, the decay of the background field value would be exponentially fast only for $m_{\rm eff}\gtrsim 0.2H_{\rm inf}$ within the total time scale of inflation. If we assume $\xi_{\rm \sigma} R \gg m_\sigma^2$, i.e. $m_\sigma^2/H_{\rm inf}^2 \ll 12\xi_\sigma$, this means that the non-minimal coupling has to take a value
\begin{equation}
\label{xi_limit}
0.0026 \leq \xi_\sigma \leq \frac{3}{16}\,,
\end{equation}
where the lower limit is given by $N=N(V_{\rm min})$, see Eq. \eqref{Nlimit_Vmin}, and the upper limit by the criterion $m_{\rm eff} < 3/2 H_{\rm inf}$.

If, after inflation, the field was locally displaced from its vacuum state, it started to oscillate about the minimum of its potential. This happened when $H(t_{\rm osc})\sim m_{\sigma}$, where $t_{\rm osc}>t_{\rm end}$. Because the potential is quadratic about the minimum, one can show that on average the energy density of the field scales down as that of cold DM, $\rho_\sigma\propto a^{-3}$. The contribution of fluctuations to the total DM abundance at the present day is then given by \cite{Tenkanen:2019aij}
\begin{equation}
\label{DMabundance}
\frac{\Omega_\sigma h^2}{0.12} = 3.5\times 10^{17}g_*^{-1/4}(t_{\rm osc})\left(\frac{\sigma_*}{M_{\rm P}}\right)^2\sqrt{\frac{m_\sigma}{{\rm GeV}}} ,
\end{equation}
where $\sigma_*$ is the local DM density and $g(t_{\rm osc})$ denotes the effective number of degrees of freedom at the time when the displaced field started oscillating about its minimum. By taking $\sigma_*=\sqrt{\langle\delta\sigma^2 \rangle}$ where the typical fluctuation is given by Eq. \eqref{sigma_fluctuations}, one obtains the typical contribution to the DM abundance. The result shows that there exist combinations of $H_{\rm inf}$, $m_\sigma$, and $\xi_\sigma$ for which the fluctuations constitute all DM. 

However, if the scale of inflation is constrained to small values and the total duration of inflation is limited, $N \lesssim 100$, this has serious consequences on this type of DM models. By first assuming that the non-minimal coupling to gravity is subdominant during inflation, assuming that the fluctuations indeed accumulate (i.e. $3/(2N) < m_\sigma^2/H_{\rm inf}^2 < 9/4$) and requiring $\Omega_\sigma h^2=0.12$, we obtain
\begin{equation}
\label{mH_relation}
\frac{m_\sigma}{\rm GeV} \simeq 8\times 10^{-15}\left(\frac{H_{\rm inf}}{\rm GeV}\right)^{8/3}
< 2\times 10^{-16}\left(\frac{r}{10^{-30}}\right)^{4/3}\,,
\end{equation}
where we assumed $g_*(t_{\rm osc})\simeq 100$ and the last inequality assumes Eq. \eqref{H_limit}. Thus, for $r=r_{\rm max} = 2\times 10^{-31}$ (see Eq. \eqref{rmax}), we obtain 
\begin{equation}
m_\sigma < 0.02\, \mu{\rm eV}.
\end{equation}
However, this mass is orders of magnitude smaller than the value of $H_{\rm inf}$, and hence in conflict with our assumptions, namely that $m_\sigma^2/H_{\rm inf}^2 > 3/(2N)$. Thus, this scenario is incompatible with the TCC and, more generally, with scenarios where the scale of inflation is constrained to small values and the total duration of inflation is limited. As is evident from Eq. \eqref{mH_relation}, smaller values of $H_{\rm inf}$ do not change this conclusion.

Let us then assume that the non-minimal coupling to gravity controls the accumulation of quantum fluctuations during inflation (i.e. $m_\sigma^2/H_{\rm inf}^2 \ll 12\xi_\sigma < 9/4$) but that it vanishes after inflation (because $R=0$ in a radiation-dominated Universe). Requiring again $\Omega_\sigma h^2=0.12$ and that the value of $\xi_\sigma$ is within the limits given by Eq. \eqref{xi_limit}, we obtain
\begin{equation}
\label{mHxi_relation}
\frac{m_\sigma}{\rm GeV} \simeq 10^{44}\left(\frac{H_{\rm inf}}{\rm GeV}\right)^{-4}\frac{\xi_\sigma^2}{\left(1-e^{-8\xi_\sigma N}\right)^2}\,.
\end{equation}
Again, the scenario is clearly incompatible with the TCC, which in this case requires $m_\sigma \ll H_{\rm inf}\lesssim 0.1$ GeV. 

The result \eqref{mHxi_relation} also shows that scenarios where the spectator field couples to some other fields during inflation which generate it an effective mass $m_{\rm eff}^2\sim c H_{\rm inf}^2$, where $c$ is a number not much smaller than unity (as in Ref. \cite{Bertolami:2016ywc}), are not compatible with the TCC or, again, more generally with scenarios where the scale of inflation and its total duration are limited. A similar result also applies to the case where the spectator field is strongly self-interacting. As shown in Ref. \cite{Markkanen:2018gcw}, in this case the allowed parameter space does not extend to values smaller than $H_{\rm inf}\simeq 10^8$ GeV, at least if the field which exhibits large self-interactions is required to constitute all DM. Therefore, also this scenario is incompatible with the assumptions presented above.

The analysis conducted above can be easily modified to accommodate also other post-inflationary cosmological histories or models where the DM sector has a richer structure. For example, one can ask if the spectator field could still decay into DM, thus mediating the accumulated energy density in inflationary fluctuations into DM. However, this seems difficult because for small $H_{\rm inf}$ the energy density stored in the quantum fluctuations is always very small. As a proof of principle, one can consider a scenario where the scalar field decays after inflation into relativistic hidden sector particles $\psi$ which never entered into thermal equilibrium with radiation. In that case, their contribution to the present DM abundance is given by \cite{Tenkanen:2019aij}
\begin{equation}
\frac{\Omega_\psi h^2}{0.12} \simeq 10^{8}\left(\frac{m_\sigma}{\Gamma_\sigma}\right)^{3/8}\left(\frac{\sigma_*}{M_{\rm P}}\right)^{3/2}\left(\frac{m_\psi}{\rm GeV}\right) ,
\end{equation}
where $\Gamma_\sigma$ is the decay width of $\sigma$. However, if we require $\sigma_*^2\simeq H_{\rm inf}^4/m_\sigma^2 \ll M_{\rm P}^2$ and $m_\psi < m_\sigma \sim H_{\rm inf}$, the required value of $\Gamma_\sigma$ becomes so small that the $\sigma$ field would not have decayed by the present day, $1/\Gamma_\sigma \gg 1/H_0$, and thus invalidating our assumption of its decay into $\psi$ particles. We leave considerations of more elaborate scenarios for future work.

Finally, we point out that while in this paper we have mainly been interested in amplification of spectator field fluctuations and its consequences on DM, the fact that a small $H_{\rm inf}$ forces quantum fluctuations to be small may have implications also for other scenarios containing extra scalar fields. An example is the curvaton model \cite{Enqvist:2001zp,Lyth:2001nq,Moroi:2001ct}, where a scalar field is subdominant during inflation but later becomes energetically important, sourcing all or part of the observed curvature perturbation. It would be interesting to assess to what extent this type of scenarios are compatible with the combination of low inflationary scale and small duration of inflation and observations.

%%%%%%%%%%%%%%%%%%%%%%%%%%%%%%%%%%%%%%%%%%%%%%%%%%%%%%%%%%%%%%%%%%%%%%%%%%%%%%%%%%%%%%%%%%

\section{Conclusions}
\label{conclusions}

In this paper, we have presented a model of inflation which exhibits a very low scale of inflation, $H_{\rm inf}\lesssim 1$ GeV, and thus satisfies the Trans-Planckian Conjecture, which has recently been introduced in the literature. In addition to being compatible with the TCC, the model, namely the one where the action contains a scalar field with a simple polynomial potential and an $R^2$ term in Palatini gravity, is also in perfect agreement with observational data. 

We also showed that the combination of a low inflationary scale and small duration of inflation the TCC anticipates is generically catastrophic for models where the DM density was generated by amplification of quantum fluctuations during inflation. This is because a low inflationary scale and small total duration of inflation prevent accumulation of large quantum fluctuations during inflation, which suppresses the energy density stored in spectator field fluctuations. As a result, if the TCC or, more generally, our assumptions of the low inflationary scale and its small total duration hold, such quantum fluctuations can never account for all of the observed DM abundance.

%%%%%%%%%%%%%%%%%%%%%%%%%%%%%%%%%%%%%%%%%%%%%%%%%%%%%%%%%%%%%%%%%%%%%%%%%%%%%%%%%%%%%%%%%%

\section*{Acknowledgements}
I thank Alan Guth, Eemeli Tomberg, and Vincent Vennin for correspondence and discussions, as well as the anonymous referees for constructive feedback, and acknowledge the Simons foundation for funding.

%%%%%%%%%%%%%%%%%%%%%%%%%%%%%%%%%%%%%%%%%%%%%%%%%%%%%%%%%%%%%%%%%%%%%%%%%%%%%%%%%%%%%%%%%%

\bibliography{TCC}

\begin{thebibliography}{85}%
\makeatletter
\providecommand \@ifxundefined [1]{%
 \@ifx{#1\undefined}
}%
\providecommand \@ifnum [1]{%
 \ifnum #1\expandafter \@firstoftwo
 \else \expandafter \@secondoftwo
 \fi
}%
\providecommand \@ifx [1]{%
 \ifx #1\expandafter \@firstoftwo
 \else \expandafter \@secondoftwo
 \fi
}%
\providecommand \natexlab [1]{#1}%
\providecommand \enquote  [1]{``#1''}%
\providecommand \bibnamefont  [1]{#1}%
\providecommand \bibfnamefont [1]{#1}%
\providecommand \citenamefont [1]{#1}%
\providecommand \href@noop [0]{\@secondoftwo}%
\providecommand \href [0]{\begingroup \@sanitize@url \@href}%
\providecommand \@href[1]{\@@startlink{#1}\@@href}%
\providecommand \@@href[1]{\endgroup#1\@@endlink}%
\providecommand \@sanitize@url [0]{\catcode `\\12\catcode `\$12\catcode
  `\&12\catcode `\#12\catcode `\^12\catcode `\_12\catcode `\%12\relax}%
\providecommand \@@startlink[1]{}%
\providecommand \@@endlink[0]{}%
\providecommand \url  [0]{\begingroup\@sanitize@url \@url }%
\providecommand \@url [1]{\endgroup\@href {#1}{\urlprefix }}%
\providecommand \urlprefix  [0]{URL }%
\providecommand \Eprint [0]{\href }%
\providecommand \doibase [0]{http://dx.doi.org/}%
\providecommand \selectlanguage [0]{\@gobble}%
\providecommand \bibinfo  [0]{\@secondoftwo}%
\providecommand \bibfield  [0]{\@secondoftwo}%
\providecommand \translation [1]{[#1]}%
\providecommand \BibitemOpen [0]{}%
\providecommand \bibitemStop [0]{}%
\providecommand \bibitemNoStop [0]{.\EOS\space}%
\providecommand \EOS [0]{\spacefactor3000\relax}%
\providecommand \BibitemShut  [1]{\csname bibitem#1\endcsname}%
\let\auto@bib@innerbib\@empty
%</preamble>
\bibitem [{\citenamefont {Starobinsky}(1980)}]{Starobinsky:1980te}%
  \BibitemOpen
  \bibfield  {author} {\bibinfo {author} {\bibfnamefont {A.~A.}\ \bibnamefont
  {Starobinsky}},\ }\href {\doibase 10.1016/0370-2693(80)90670-X} {\bibfield
  {journal} {\bibinfo  {journal} {Phys. Lett.}\ }\textbf {\bibinfo {volume}
  {B91}},\ \bibinfo {pages} {99} (\bibinfo {year} {1980})}\BibitemShut
  {NoStop}%
%%CITATION = PHLTA,B91,99;%%
\bibitem [{\citenamefont {Sato}(1981)}]{Sato:1980yn}%
  \BibitemOpen
  \bibfield  {author} {\bibinfo {author} {\bibfnamefont {K.}~\bibnamefont
  {Sato}},\ }\href@noop {} {\bibfield  {journal} {\bibinfo  {journal}
  {Mon.Not.Roy.Astron.Soc.}\ }\textbf {\bibinfo {volume} {195}},\ \bibinfo
  {pages} {467} (\bibinfo {year} {1981})}\BibitemShut {NoStop}%
%%CITATION = MNRAA,195,467;%%
\bibitem [{\citenamefont {Guth}(1981)}]{Guth:1980zm}%
  \BibitemOpen
  \bibfield  {author} {\bibinfo {author} {\bibfnamefont {A.~H.}\ \bibnamefont
  {Guth}},\ }\href {\doibase 10.1103/PhysRevD.23.347} {\bibfield  {journal}
  {\bibinfo  {journal} {Phys.Rev.}\ }\textbf {\bibinfo {volume} {D23}},\
  \bibinfo {pages} {347} (\bibinfo {year} {1981})}\BibitemShut {NoStop}%
%%CITATION = PHRVA,D23,347;%%
\bibitem [{\citenamefont {Linde}(1982)}]{Linde:1981mu}%
  \BibitemOpen
  \bibfield  {author} {\bibinfo {author} {\bibfnamefont {A.~D.}\ \bibnamefont
  {Linde}},\ }\href {\doibase 10.1016/0370-2693(82)91219-9} {\bibfield
  {journal} {\bibinfo  {journal} {Phys.Lett.}\ }\textbf {\bibinfo {volume}
  {B108}},\ \bibinfo {pages} {389} (\bibinfo {year} {1982})}\BibitemShut
  {NoStop}%
%%CITATION = PHLTA,B108,389;%%
\bibitem [{\citenamefont {Albrecht}\ and\ \citenamefont
  {Steinhardt}(1982)}]{Albrecht:1982wi}%
  \BibitemOpen
  \bibfield  {author} {\bibinfo {author} {\bibfnamefont {A.}~\bibnamefont
  {Albrecht}}\ and\ \bibinfo {author} {\bibfnamefont {P.~J.}\ \bibnamefont
  {Steinhardt}},\ }\href {\doibase 10.1103/PhysRevLett.48.1220} {\bibfield
  {journal} {\bibinfo  {journal} {Phys.Rev.Lett.}\ }\textbf {\bibinfo {volume}
  {48}},\ \bibinfo {pages} {1220} (\bibinfo {year} {1982})}\BibitemShut
  {NoStop}%
%%CITATION = PRLTA,48,1220;%%
\bibitem [{\citenamefont {Linde}(1983)}]{Linde:1983gd}%
  \BibitemOpen
  \bibfield  {author} {\bibinfo {author} {\bibfnamefont {A.~D.}\ \bibnamefont
  {Linde}},\ }\href {\doibase 10.1016/0370-2693(83)90837-7} {\bibfield
  {journal} {\bibinfo  {journal} {Phys.Lett.}\ }\textbf {\bibinfo {volume}
  {B129}},\ \bibinfo {pages} {177} (\bibinfo {year} {1983})}\BibitemShut
  {NoStop}%
%%CITATION = PHLTA,B129,177;%%
\bibitem [{\citenamefont {Martin}\ \emph {et~al.}(2014)\citenamefont {Martin},
  \citenamefont {Ringeval},\ and\ \citenamefont {Vennin}}]{Martin:2013tda}%
  \BibitemOpen
  \bibfield  {author} {\bibinfo {author} {\bibfnamefont {J.}~\bibnamefont
  {Martin}}, \bibinfo {author} {\bibfnamefont {C.}~\bibnamefont {Ringeval}}, \
  and\ \bibinfo {author} {\bibfnamefont {V.}~\bibnamefont {Vennin}},\ }\href
  {\doibase 10.1016/j.dark.2014.01.003} {\bibfield  {journal} {\bibinfo
  {journal} {Phys. Dark Univ.}\ }\textbf {\bibinfo {volume} {5-6}},\ \bibinfo
  {pages} {75} (\bibinfo {year} {2014})},\ \Eprint
  {http://arxiv.org/abs/1303.3787} {arXiv:1303.3787 [astro-ph.CO]} \BibitemShut
  {NoStop}%
%%CITATION = ARXIV:1303.3787;%%
\bibitem [{\citenamefont {Akrami}\ \emph {et~al.}(2018)\citenamefont {Akrami}
  \emph {et~al.}}]{Akrami:2018odb}%
  \BibitemOpen
  \bibfield  {author} {\bibinfo {author} {\bibfnamefont {Y.}~\bibnamefont
  {Akrami}} \emph {et~al.} (\bibinfo {collaboration} {Planck}),\ }\href@noop {}
  {\  (\bibinfo {year} {2018})},\ \Eprint {http://arxiv.org/abs/1807.06211}
  {arXiv:1807.06211 [astro-ph.CO]} \BibitemShut {NoStop}%
%%CITATION = ARXIV:1807.06211;%%
\bibitem [{\citenamefont {Chowdhury}\ \emph {et~al.}(2019)\citenamefont
  {Chowdhury}, \citenamefont {Martin}, \citenamefont {Ringeval},\ and\
  \citenamefont {Vennin}}]{Chowdhury:2019otk}%
  \BibitemOpen
  \bibfield  {author} {\bibinfo {author} {\bibfnamefont {D.}~\bibnamefont
  {Chowdhury}}, \bibinfo {author} {\bibfnamefont {J.}~\bibnamefont {Martin}},
  \bibinfo {author} {\bibfnamefont {C.}~\bibnamefont {Ringeval}}, \ and\
  \bibinfo {author} {\bibfnamefont {V.}~\bibnamefont {Vennin}},\ }\href
  {\doibase 10.1103/PhysRevD.100.083537} {\bibfield  {journal} {\bibinfo
  {journal} {Phys. Rev.}\ }\textbf {\bibinfo {volume} {D100}},\ \bibinfo
  {pages} {083537} (\bibinfo {year} {2019})},\ \Eprint
  {http://arxiv.org/abs/1902.03951} {arXiv:1902.03951 [astro-ph.CO]}
  \BibitemShut {NoStop}%
%%CITATION = ARXIV:1902.03951;%%
\bibitem [{\citenamefont {Bertone}\ and\ \citenamefont
  {Hooper}(2018)}]{Bertone:2016nfn}%
  \BibitemOpen
  \bibfield  {author} {\bibinfo {author} {\bibfnamefont {G.}~\bibnamefont
  {Bertone}}\ and\ \bibinfo {author} {\bibfnamefont {D.}~\bibnamefont
  {Hooper}},\ }\href {\doibase 10.1103/RevModPhys.90.045002} {\bibfield
  {journal} {\bibinfo  {journal} {Rev. Mod. Phys.}\ }\textbf {\bibinfo {volume}
  {90}},\ \bibinfo {pages} {045002} (\bibinfo {year} {2018})},\ \Eprint
  {http://arxiv.org/abs/1605.04909} {arXiv:1605.04909 [astro-ph.CO]}
  \BibitemShut {NoStop}%
%%CITATION = ARXIV:1605.04909;%%
\bibitem [{\citenamefont {Aghanim}\ \emph {et~al.}(2018)\citenamefont {Aghanim}
  \emph {et~al.}}]{Aghanim:2018eyx}%
  \BibitemOpen
  \bibfield  {author} {\bibinfo {author} {\bibfnamefont {N.}~\bibnamefont
  {Aghanim}} \emph {et~al.} (\bibinfo {collaboration} {Planck}),\ }\href@noop
  {} {\  (\bibinfo {year} {2018})},\ \Eprint {http://arxiv.org/abs/1807.06209}
  {arXiv:1807.06209 [astro-ph.CO]} \BibitemShut {NoStop}%
%%CITATION = ARXIV:1807.06209;%%
\bibitem [{\citenamefont {Kolb}\ and\ \citenamefont
  {Turner}(1990)}]{Kolb:1990vq}%
  \BibitemOpen
  \bibfield  {author} {\bibinfo {author} {\bibfnamefont {E.~W.}\ \bibnamefont
  {Kolb}}\ and\ \bibinfo {author} {\bibfnamefont {M.~S.}\ \bibnamefont
  {Turner}},\ }\href@noop {} {\bibfield  {journal} {\bibinfo  {journal} {Front.
  Phys.}\ }\textbf {\bibinfo {volume} {69}},\ \bibinfo {pages} {1} (\bibinfo
  {year} {1990})}\BibitemShut {NoStop}%
%%CITATION = FRPHA,69,1;%%
\bibitem [{\citenamefont {McDonald}(2002)}]{McDonald:2001vt}%
  \BibitemOpen
  \bibfield  {author} {\bibinfo {author} {\bibfnamefont {J.}~\bibnamefont
  {McDonald}},\ }\href {\doibase 10.1103/PhysRevLett.88.091304} {\bibfield
  {journal} {\bibinfo  {journal} {Phys.Rev.Lett.}\ }\textbf {\bibinfo {volume}
  {88}},\ \bibinfo {pages} {091304} (\bibinfo {year} {2002})},\ \Eprint
  {http://arxiv.org/abs/hep-ph/0106249} {arXiv:hep-ph/0106249 [hep-ph]}
  \BibitemShut {NoStop}%
%%CITATION = HEP-PH/0106249;%%
\bibitem [{\citenamefont {Hall}\ \emph {et~al.}(2010)\citenamefont {Hall},
  \citenamefont {Jedamzik}, \citenamefont {March-Russell},\ and\ \citenamefont
  {West}}]{Hall:2009bx}%
  \BibitemOpen
  \bibfield  {author} {\bibinfo {author} {\bibfnamefont {L.~J.}\ \bibnamefont
  {Hall}}, \bibinfo {author} {\bibfnamefont {K.}~\bibnamefont {Jedamzik}},
  \bibinfo {author} {\bibfnamefont {J.}~\bibnamefont {March-Russell}}, \ and\
  \bibinfo {author} {\bibfnamefont {S.~M.}\ \bibnamefont {West}},\ }\href
  {\doibase 10.1007/JHEP03(2010)080} {\bibfield  {journal} {\bibinfo  {journal}
  {JHEP}\ }\textbf {\bibinfo {volume} {1003}},\ \bibinfo {pages} {080}
  (\bibinfo {year} {2010})},\ \Eprint {http://arxiv.org/abs/0911.1120}
  {arXiv:0911.1120 [hep-ph]} \BibitemShut {NoStop}%
%%CITATION = ARXIV:0911.1120;%%
\bibitem [{\citenamefont {Bernal}\ \emph {et~al.}(2017)\citenamefont {Bernal},
  \citenamefont {Heikinheimo}, \citenamefont {Tenkanen}, \citenamefont
  {Tuominen},\ and\ \citenamefont {Vaskonen}}]{Bernal:2017kxu}%
  \BibitemOpen
  \bibfield  {author} {\bibinfo {author} {\bibfnamefont {N.}~\bibnamefont
  {Bernal}}, \bibinfo {author} {\bibfnamefont {M.}~\bibnamefont {Heikinheimo}},
  \bibinfo {author} {\bibfnamefont {T.}~\bibnamefont {Tenkanen}}, \bibinfo
  {author} {\bibfnamefont {K.}~\bibnamefont {Tuominen}}, \ and\ \bibinfo
  {author} {\bibfnamefont {V.}~\bibnamefont {Vaskonen}},\ }\href {\doibase
  10.1142/S0217751X1730023X} {\bibfield  {journal} {\bibinfo  {journal} {Int.
  J. Mod. Phys.}\ }\textbf {\bibinfo {volume} {A32}},\ \bibinfo {pages}
  {1730023} (\bibinfo {year} {2017})},\ \Eprint
  {http://arxiv.org/abs/1706.07442} {arXiv:1706.07442 [hep-ph]} \BibitemShut
  {NoStop}%
%%CITATION = ARXIV:1706.07442;%%
\bibitem [{\citenamefont {Ford}(1987)}]{Ford:1986sy}%
  \BibitemOpen
  \bibfield  {author} {\bibinfo {author} {\bibfnamefont {L.~H.}\ \bibnamefont
  {Ford}},\ }\href {\doibase 10.1103/PhysRevD.35.2955} {\bibfield  {journal}
  {\bibinfo  {journal} {Phys. Rev.}\ }\textbf {\bibinfo {volume} {D35}},\
  \bibinfo {pages} {2955} (\bibinfo {year} {1987})}\BibitemShut {NoStop}%
%%CITATION = PHRVA,D35,2955;%%
\bibitem [{\citenamefont {Turner}\ and\ \citenamefont
  {Widrow}(1988)}]{Turner:1987vd}%
  \BibitemOpen
  \bibfield  {author} {\bibinfo {author} {\bibfnamefont {M.~S.}\ \bibnamefont
  {Turner}}\ and\ \bibinfo {author} {\bibfnamefont {L.~M.}\ \bibnamefont
  {Widrow}},\ }\href {\doibase 10.1103/PhysRevD.37.3428} {\bibfield  {journal}
  {\bibinfo  {journal} {Phys. Rev.}\ }\textbf {\bibinfo {volume} {D37}},\
  \bibinfo {pages} {3428} (\bibinfo {year} {1988})}\BibitemShut {NoStop}%
%%CITATION = PHRVA,D37,3428;%%
\bibitem [{\citenamefont {Kolb}\ \emph {et~al.}(1999)\citenamefont {Kolb},
  \citenamefont {Chung},\ and\ \citenamefont {Riotto}}]{Kolb:1998ki}%
  \BibitemOpen
  \bibfield  {author} {\bibinfo {author} {\bibfnamefont {E.~W.}\ \bibnamefont
  {Kolb}}, \bibinfo {author} {\bibfnamefont {D.~J.~H.}\ \bibnamefont {Chung}},
  \ and\ \bibinfo {author} {\bibfnamefont {A.}~\bibnamefont {Riotto}},\ }\href
  {\doibase 10.1063/1.59655} {\bibfield  {journal} {\bibinfo  {journal} {AIP
  Conf. Proc.}\ }\textbf {\bibinfo {volume} {484}},\ \bibinfo {pages} {91}
  (\bibinfo {year} {1999})},\ \bibinfo {note} {[592(1999)]},\ \Eprint
  {http://arxiv.org/abs/hep-ph/9810361} {arXiv:hep-ph/9810361 [hep-ph]}
  \BibitemShut {NoStop}%
%%CITATION = HEP-PH/9810361;%%
\bibitem [{\citenamefont {Chung}\ \emph {et~al.}(1999)\citenamefont {Chung},
  \citenamefont {Kolb},\ and\ \citenamefont {Riotto}}]{Chung:1998zb}%
  \BibitemOpen
  \bibfield  {author} {\bibinfo {author} {\bibfnamefont {D.~J.~H.}\
  \bibnamefont {Chung}}, \bibinfo {author} {\bibfnamefont {E.~W.}\ \bibnamefont
  {Kolb}}, \ and\ \bibinfo {author} {\bibfnamefont {A.}~\bibnamefont
  {Riotto}},\ }\href {\doibase 10.1103/PhysRevD.59.023501} {\bibfield
  {journal} {\bibinfo  {journal} {Phys. Rev.}\ }\textbf {\bibinfo {volume}
  {D59}},\ \bibinfo {pages} {023501} (\bibinfo {year} {1999})},\ \Eprint
  {http://arxiv.org/abs/hep-ph/9802238} {arXiv:hep-ph/9802238 [hep-ph]}
  \BibitemShut {NoStop}%
%%CITATION = HEP-PH/9802238;%%
\bibitem [{\citenamefont {Peebles}\ and\ \citenamefont
  {Vilenkin}(1999)}]{Peebles:1999fz}%
  \BibitemOpen
  \bibfield  {author} {\bibinfo {author} {\bibfnamefont {P.~J.~E.}\
  \bibnamefont {Peebles}}\ and\ \bibinfo {author} {\bibfnamefont
  {A.}~\bibnamefont {Vilenkin}},\ }\href {\doibase 10.1103/PhysRevD.60.103506}
  {\bibfield  {journal} {\bibinfo  {journal} {Phys. Rev.}\ }\textbf {\bibinfo
  {volume} {D60}},\ \bibinfo {pages} {103506} (\bibinfo {year} {1999})},\
  \Eprint {http://arxiv.org/abs/astro-ph/9904396} {arXiv:astro-ph/9904396
  [astro-ph]} \BibitemShut {NoStop}%
%%CITATION = ASTRO-PH/9904396;%%
\bibitem [{\citenamefont {Nurmi}\ \emph {et~al.}(2015)\citenamefont {Nurmi},
  \citenamefont {Tenkanen},\ and\ \citenamefont {Tuominen}}]{Nurmi:2015ema}%
  \BibitemOpen
  \bibfield  {author} {\bibinfo {author} {\bibfnamefont {S.}~\bibnamefont
  {Nurmi}}, \bibinfo {author} {\bibfnamefont {T.}~\bibnamefont {Tenkanen}}, \
  and\ \bibinfo {author} {\bibfnamefont {K.}~\bibnamefont {Tuominen}},\ }\href
  {\doibase 10.1088/1475-7516/2015/11/001} {\bibfield  {journal} {\bibinfo
  {journal} {JCAP}\ }\textbf {\bibinfo {volume} {1511}},\ \bibinfo {pages}
  {001} (\bibinfo {year} {2015})},\ \Eprint {http://arxiv.org/abs/1506.04048}
  {arXiv:1506.04048 [astro-ph.CO]} \BibitemShut {NoStop}%
%%CITATION = ARXIV:1506.04048;%%
\bibitem [{\citenamefont {Graham}\ \emph {et~al.}(2016)\citenamefont {Graham},
  \citenamefont {Mardon},\ and\ \citenamefont {Rajendran}}]{Graham:2015rva}%
  \BibitemOpen
  \bibfield  {author} {\bibinfo {author} {\bibfnamefont {P.~W.}\ \bibnamefont
  {Graham}}, \bibinfo {author} {\bibfnamefont {J.}~\bibnamefont {Mardon}}, \
  and\ \bibinfo {author} {\bibfnamefont {S.}~\bibnamefont {Rajendran}},\ }\href
  {\doibase 10.1103/PhysRevD.93.103520} {\bibfield  {journal} {\bibinfo
  {journal} {Phys. Rev.}\ }\textbf {\bibinfo {volume} {D93}},\ \bibinfo {pages}
  {103520} (\bibinfo {year} {2016})},\ \Eprint
  {http://arxiv.org/abs/1504.02102} {arXiv:1504.02102 [hep-ph]} \BibitemShut
  {NoStop}%
%%CITATION = ARXIV:1504.02102;%%
\bibitem [{\citenamefont {Markkanen}\ and\ \citenamefont
  {Nurmi}(2017)}]{Markkanen:2015xuw}%
  \BibitemOpen
  \bibfield  {author} {\bibinfo {author} {\bibfnamefont {T.}~\bibnamefont
  {Markkanen}}\ and\ \bibinfo {author} {\bibfnamefont {S.}~\bibnamefont
  {Nurmi}},\ }\href {\doibase 10.1088/1475-7516/2017/02/008} {\bibfield
  {journal} {\bibinfo  {journal} {JCAP}\ }\textbf {\bibinfo {volume} {1702}},\
  \bibinfo {pages} {008} (\bibinfo {year} {2017})},\ \Eprint
  {http://arxiv.org/abs/1512.07288} {arXiv:1512.07288 [astro-ph.CO]}
  \BibitemShut {NoStop}%
%%CITATION = ARXIV:1512.07288;%%
\bibitem [{\citenamefont {Bertolami}\ \emph {et~al.}(2016)\citenamefont
  {Bertolami}, \citenamefont {Cosme},\ and\ \citenamefont
  {Rosa}}]{Bertolami:2016ywc}%
  \BibitemOpen
  \bibfield  {author} {\bibinfo {author} {\bibfnamefont {O.}~\bibnamefont
  {Bertolami}}, \bibinfo {author} {\bibfnamefont {C.}~\bibnamefont {Cosme}}, \
  and\ \bibinfo {author} {\bibfnamefont {J.~G.}\ \bibnamefont {Rosa}},\ }\href
  {\doibase 10.1016/j.physletb.2016.05.047} {\bibfield  {journal} {\bibinfo
  {journal} {Phys. Lett.}\ }\textbf {\bibinfo {volume} {B759}},\ \bibinfo
  {pages} {1} (\bibinfo {year} {2016})},\ \Eprint
  {http://arxiv.org/abs/1603.06242} {arXiv:1603.06242 [hep-ph]} \BibitemShut
  {NoStop}%
%%CITATION = ARXIV:1603.06242;%%
\bibitem [{\citenamefont {Cosme}\ \emph
  {et~al.}(2018{\natexlab{a}})\citenamefont {Cosme}, \citenamefont {Rosa},\
  and\ \citenamefont {Bertolami}}]{Cosme:2017cxk}%
  \BibitemOpen
  \bibfield  {author} {\bibinfo {author} {\bibfnamefont {C.}~\bibnamefont
  {Cosme}}, \bibinfo {author} {\bibfnamefont {J.~G.}\ \bibnamefont {Rosa}}, \
  and\ \bibinfo {author} {\bibfnamefont {O.}~\bibnamefont {Bertolami}},\ }\href
  {\doibase 10.1016/j.physletb.2018.04.062} {\bibfield  {journal} {\bibinfo
  {journal} {Phys. Lett.}\ }\textbf {\bibinfo {volume} {B781}},\ \bibinfo
  {pages} {639} (\bibinfo {year} {2018}{\natexlab{a}})},\ \Eprint
  {http://arxiv.org/abs/1709.09674} {arXiv:1709.09674 [hep-ph]} \BibitemShut
  {NoStop}%
%%CITATION = ARXIV:1709.09674;%%
\bibitem [{\citenamefont {Alonso-Álvarez}\ and\ \citenamefont
  {Jaeckel}(2018)}]{Alonso-Alvarez:2018tus}%
  \BibitemOpen
  \bibfield  {author} {\bibinfo {author} {\bibfnamefont {G.}~\bibnamefont
  {Alonso-Álvarez}}\ and\ \bibinfo {author} {\bibfnamefont {J.}~\bibnamefont
  {Jaeckel}},\ }\href {\doibase 10.1088/1475-7516/2018/10/022} {\bibfield
  {journal} {\bibinfo  {journal} {JCAP}\ }\textbf {\bibinfo {volume} {1810}},\
  \bibinfo {pages} {022} (\bibinfo {year} {2018})},\ \Eprint
  {http://arxiv.org/abs/1807.09785} {arXiv:1807.09785 [hep-ph]} \BibitemShut
  {NoStop}%
%%CITATION = ARXIV:1807.09785;%%
\bibitem [{\citenamefont {Fairbairn}\ \emph {et~al.}(2019)\citenamefont
  {Fairbairn}, \citenamefont {Kainulainen}, \citenamefont {Markkanen},\ and\
  \citenamefont {Nurmi}}]{Fairbairn:2018bsw}%
  \BibitemOpen
  \bibfield  {author} {\bibinfo {author} {\bibfnamefont {M.}~\bibnamefont
  {Fairbairn}}, \bibinfo {author} {\bibfnamefont {K.}~\bibnamefont
  {Kainulainen}}, \bibinfo {author} {\bibfnamefont {T.}~\bibnamefont
  {Markkanen}}, \ and\ \bibinfo {author} {\bibfnamefont {S.}~\bibnamefont
  {Nurmi}},\ }\href {\doibase 10.1088/1475-7516/2019/04/005} {\bibfield
  {journal} {\bibinfo  {journal} {JCAP}\ }\textbf {\bibinfo {volume} {1904}},\
  \bibinfo {pages} {005} (\bibinfo {year} {2019})},\ \Eprint
  {http://arxiv.org/abs/1808.08236} {arXiv:1808.08236 [astro-ph.CO]}
  \BibitemShut {NoStop}%
%%CITATION = ARXIV:1808.08236;%%
\bibitem [{\citenamefont {Markkanen}\ \emph {et~al.}(2018)\citenamefont
  {Markkanen}, \citenamefont {Rajantie},\ and\ \citenamefont
  {Tenkanen}}]{Markkanen:2018gcw}%
  \BibitemOpen
  \bibfield  {author} {\bibinfo {author} {\bibfnamefont {T.}~\bibnamefont
  {Markkanen}}, \bibinfo {author} {\bibfnamefont {A.}~\bibnamefont {Rajantie}},
  \ and\ \bibinfo {author} {\bibfnamefont {T.}~\bibnamefont {Tenkanen}},\
  }\href {\doibase 10.1103/PhysRevD.98.123532} {\bibfield  {journal} {\bibinfo
  {journal} {Phys. Rev.}\ }\textbf {\bibinfo {volume} {D98}},\ \bibinfo {pages}
  {123532} (\bibinfo {year} {2018})},\ \Eprint
  {http://arxiv.org/abs/1811.02586} {arXiv:1811.02586 [astro-ph.CO]}
  \BibitemShut {NoStop}%
%%CITATION = ARXIV:1811.02586;%%
\bibitem [{\citenamefont {Tenkanen}(2019{\natexlab{a}})}]{Tenkanen:2019aij}%
  \BibitemOpen
  \bibfield  {author} {\bibinfo {author} {\bibfnamefont {T.}~\bibnamefont
  {Tenkanen}},\ }\href {\doibase 10.1103/PhysRevLett.123.061302} {\bibfield
  {journal} {\bibinfo  {journal} {Phys. Rev. Lett.}\ }\textbf {\bibinfo
  {volume} {123}},\ \bibinfo {pages} {061302} (\bibinfo {year}
  {2019}{\natexlab{a}})},\ \Eprint {http://arxiv.org/abs/1905.01214}
  {arXiv:1905.01214 [astro-ph.CO]} \BibitemShut {NoStop}%
%%CITATION = ARXIV:1905.01214;%%
\bibitem [{\citenamefont {Alonso-Álvarez}\ \emph {et~al.}(2020)\citenamefont
  {Alonso-Álvarez}, \citenamefont {Jaeckel},\ and\ \citenamefont
  {Hugle}}]{AlonsoAlvarez:2019cgw}%
  \BibitemOpen
  \bibfield  {author} {\bibinfo {author} {\bibfnamefont {G.}~\bibnamefont
  {Alonso-Álvarez}}, \bibinfo {author} {\bibfnamefont {J.}~\bibnamefont
  {Jaeckel}}, \ and\ \bibinfo {author} {\bibfnamefont {T.}~\bibnamefont
  {Hugle}},\ }\href {\doibase 10.1088/1475-7516/2020/02/014} {\bibfield
  {journal} {\bibinfo  {journal} {JCAP}\ }\textbf {\bibinfo {volume} {2002}},\
  \bibinfo {pages} {014} (\bibinfo {year} {2020})},\ \Eprint
  {http://arxiv.org/abs/1905.09836} {arXiv:1905.09836 [hep-ph]} \BibitemShut
  {NoStop}%
%%CITATION = ARXIV:1905.09836;%%
\bibitem [{\citenamefont {Amendola}\ \emph {et~al.}(2018)\citenamefont
  {Amendola} \emph {et~al.}}]{Amendola:2016saw}%
  \BibitemOpen
  \bibfield  {author} {\bibinfo {author} {\bibfnamefont {L.}~\bibnamefont
  {Amendola}} \emph {et~al.},\ }\href {\doibase 10.1007/s41114-017-0010-3}
  {\bibfield  {journal} {\bibinfo  {journal} {Living Rev. Rel.}\ }\textbf
  {\bibinfo {volume} {21}},\ \bibinfo {pages} {2} (\bibinfo {year} {2018})},\
  \Eprint {http://arxiv.org/abs/1606.00180} {arXiv:1606.00180 [astro-ph.CO]}
  \BibitemShut {NoStop}%
%%CITATION = ARXIV:1606.00180;%%
\bibitem [{\citenamefont {Martin}\ and\ \citenamefont
  {Brandenberger}(2001)}]{Martin:2000xs}%
  \BibitemOpen
  \bibfield  {author} {\bibinfo {author} {\bibfnamefont {J.}~\bibnamefont
  {Martin}}\ and\ \bibinfo {author} {\bibfnamefont {R.~H.}\ \bibnamefont
  {Brandenberger}},\ }\href {\doibase 10.1103/PhysRevD.63.123501} {\bibfield
  {journal} {\bibinfo  {journal} {Phys. Rev.}\ }\textbf {\bibinfo {volume}
  {D63}},\ \bibinfo {pages} {123501} (\bibinfo {year} {2001})},\ \Eprint
  {http://arxiv.org/abs/hep-th/0005209} {arXiv:hep-th/0005209 [hep-th]}
  \BibitemShut {NoStop}%
%%CITATION = HEP-TH/0005209;%%
\bibitem [{\citenamefont {Brandenberger}\ and\ \citenamefont
  {Martin}(2001)}]{Brandenberger:2000wr}%
  \BibitemOpen
  \bibfield  {author} {\bibinfo {author} {\bibfnamefont {R.~H.}\ \bibnamefont
  {Brandenberger}}\ and\ \bibinfo {author} {\bibfnamefont {J.}~\bibnamefont
  {Martin}},\ }\href {\doibase 10.1142/S0217732301004170} {\bibfield  {journal}
  {\bibinfo  {journal} {Mod. Phys. Lett.}\ }\textbf {\bibinfo {volume} {A16}},\
  \bibinfo {pages} {999} (\bibinfo {year} {2001})},\ \Eprint
  {http://arxiv.org/abs/astro-ph/0005432} {arXiv:astro-ph/0005432 [astro-ph]}
  \BibitemShut {NoStop}%
%%CITATION = ASTRO-PH/0005432;%%
\bibitem [{\citenamefont {Brandenberger}\ and\ \citenamefont
  {Martin}(2013)}]{Brandenberger:2012aj}%
  \BibitemOpen
  \bibfield  {author} {\bibinfo {author} {\bibfnamefont {R.~H.}\ \bibnamefont
  {Brandenberger}}\ and\ \bibinfo {author} {\bibfnamefont {J.}~\bibnamefont
  {Martin}},\ }\href {\doibase 10.1088/0264-9381/30/11/113001} {\bibfield
  {journal} {\bibinfo  {journal} {Class. Quant. Grav.}\ }\textbf {\bibinfo
  {volume} {30}},\ \bibinfo {pages} {113001} (\bibinfo {year} {2013})},\
  \Eprint {http://arxiv.org/abs/1211.6753} {arXiv:1211.6753 [astro-ph.CO]}
  \BibitemShut {NoStop}%
%%CITATION = ARXIV:1211.6753;%%
\bibitem [{\citenamefont {Bedroya}\ and\ \citenamefont
  {Vafa}(2019)}]{Bedroya:2019snp}%
  \BibitemOpen
  \bibfield  {author} {\bibinfo {author} {\bibfnamefont {A.}~\bibnamefont
  {Bedroya}}\ and\ \bibinfo {author} {\bibfnamefont {C.}~\bibnamefont {Vafa}},\
  }\href@noop {} {\  (\bibinfo {year} {2019})},\ \Eprint
  {http://arxiv.org/abs/1909.11063} {arXiv:1909.11063 [hep-th]} \BibitemShut
  {NoStop}%
%%CITATION = ARXIV:1909.11063;%%
\bibitem [{\citenamefont {Ooguri}\ and\ \citenamefont
  {Vafa}(2007)}]{Ooguri:2006in}%
  \BibitemOpen
  \bibfield  {author} {\bibinfo {author} {\bibfnamefont {H.}~\bibnamefont
  {Ooguri}}\ and\ \bibinfo {author} {\bibfnamefont {C.}~\bibnamefont {Vafa}},\
  }\href {\doibase 10.1016/j.nuclphysb.2006.10.033} {\bibfield  {journal}
  {\bibinfo  {journal} {Nucl. Phys.}\ }\textbf {\bibinfo {volume} {B766}},\
  \bibinfo {pages} {21} (\bibinfo {year} {2007})},\ \Eprint
  {http://arxiv.org/abs/hep-th/0605264} {arXiv:hep-th/0605264 [hep-th]}
  \BibitemShut {NoStop}%
%%CITATION = HEP-TH/0605264;%%
\bibitem [{\citenamefont {Obied}\ \emph {et~al.}(2018)\citenamefont {Obied},
  \citenamefont {Ooguri}, \citenamefont {Spodyneiko},\ and\ \citenamefont
  {Vafa}}]{Obied:2018sgi}%
  \BibitemOpen
  \bibfield  {author} {\bibinfo {author} {\bibfnamefont {G.}~\bibnamefont
  {Obied}}, \bibinfo {author} {\bibfnamefont {H.}~\bibnamefont {Ooguri}},
  \bibinfo {author} {\bibfnamefont {L.}~\bibnamefont {Spodyneiko}}, \ and\
  \bibinfo {author} {\bibfnamefont {C.}~\bibnamefont {Vafa}},\ }\href@noop {}
  {\  (\bibinfo {year} {2018})},\ \Eprint {http://arxiv.org/abs/1806.08362}
  {arXiv:1806.08362 [hep-th]} \BibitemShut {NoStop}%
%%CITATION = ARXIV:1806.08362;%%
\bibitem [{\citenamefont {Palti}(2019)}]{Palti:2019pca}%
  \BibitemOpen
  \bibfield  {author} {\bibinfo {author} {\bibfnamefont {E.}~\bibnamefont
  {Palti}},\ }\href {\doibase 10.1002/prop.201900037} {\bibfield  {journal}
  {\bibinfo  {journal} {Fortsch. Phys.}\ }\textbf {\bibinfo {volume} {67}},\
  \bibinfo {pages} {1900037} (\bibinfo {year} {2019})},\ \Eprint
  {http://arxiv.org/abs/1903.06239} {arXiv:1903.06239 [hep-th]} \BibitemShut
  {NoStop}%
%%CITATION = ARXIV:1903.06239;%%
\bibitem [{\citenamefont {Bedroya}\ \emph {et~al.}(2019)\citenamefont
  {Bedroya}, \citenamefont {Brandenberger}, \citenamefont {Loverde},\ and\
  \citenamefont {Vafa}}]{Bedroya:2019tba}%
  \BibitemOpen
  \bibfield  {author} {\bibinfo {author} {\bibfnamefont {A.}~\bibnamefont
  {Bedroya}}, \bibinfo {author} {\bibfnamefont {R.}~\bibnamefont
  {Brandenberger}}, \bibinfo {author} {\bibfnamefont {M.}~\bibnamefont
  {Loverde}}, \ and\ \bibinfo {author} {\bibfnamefont {C.}~\bibnamefont
  {Vafa}},\ }\href@noop {} {\  (\bibinfo {year} {2019})},\ \Eprint
  {http://arxiv.org/abs/1909.11106} {arXiv:1909.11106 [hep-th]} \BibitemShut
  {NoStop}%
%%CITATION = ARXIV:1909.11106;%%
\bibitem [{\citenamefont {Cai}\ and\ \citenamefont {Piao}(2019)}]{Cai:2019hge}%
  \BibitemOpen
  \bibfield  {author} {\bibinfo {author} {\bibfnamefont {Y.}~\bibnamefont
  {Cai}}\ and\ \bibinfo {author} {\bibfnamefont {Y.-S.}\ \bibnamefont {Piao}},\
  }\href@noop {} {\  (\bibinfo {year} {2019})},\ \Eprint
  {http://arxiv.org/abs/1909.12719} {arXiv:1909.12719 [gr-qc]} \BibitemShut
  {NoStop}%
%%CITATION = ARXIV:1909.12719;%%
\bibitem [{\citenamefont {Baumann}(2018)}]{Baumann:2018muz}%
  \BibitemOpen
  \bibfield  {author} {\bibinfo {author} {\bibfnamefont {D.}~\bibnamefont
  {Baumann}},\ }\href {\doibase 10.22323/1.305.0009} {\bibfield  {journal}
  {\bibinfo  {journal} {PoS}\ }\textbf {\bibinfo {volume} {TASI2017}},\
  \bibinfo {pages} {009} (\bibinfo {year} {2018})},\ \Eprint
  {http://arxiv.org/abs/1807.03098} {arXiv:1807.03098 [hep-th]} \BibitemShut
  {NoStop}%
%%CITATION = ARXIV:1807.03098;%%
\bibitem [{\citenamefont {Wu}\ \emph {et~al.}(2016)\citenamefont {Wu} \emph
  {et~al.}}]{Wu:2016hul}%
  \BibitemOpen
  \bibfield  {author} {\bibinfo {author} {\bibfnamefont {W.~L.~K.}\
  \bibnamefont {Wu}} \emph {et~al.},\ }\href {\doibase
  10.1007/s10909-015-1403-x} {\bibfield  {journal} {\bibinfo  {journal} {J.
  Low. Temp. Phys.}\ }\textbf {\bibinfo {volume} {184}},\ \bibinfo {pages}
  {765} (\bibinfo {year} {2016})},\ \Eprint {http://arxiv.org/abs/1601.00125}
  {arXiv:1601.00125 [astro-ph.IM]} \BibitemShut {NoStop}%
%%CITATION = ARXIV:1601.00125;%%
\bibitem [{\citenamefont {Matsumura}\ \emph {et~al.}(2013)\citenamefont
  {Matsumura} \emph {et~al.}}]{Matsumura:2013aja}%
  \BibitemOpen
  \bibfield  {author} {\bibinfo {author} {\bibfnamefont {T.}~\bibnamefont
  {Matsumura}} \emph {et~al.},\ }\href {\doibase 10.1007/s10909-013-0996-1} {\
  (\bibinfo {year} {2013}),\ 10.1007/s10909-013-0996-1},\ \bibinfo {note} {[J.
  Low. Temp. Phys.176,733(2014)]},\ \Eprint {http://arxiv.org/abs/1311.2847}
  {arXiv:1311.2847 [astro-ph.IM]} \BibitemShut {NoStop}%
%%CITATION = ARXIV:1311.2847;%%
\bibitem [{\citenamefont {Ade}\ \emph {et~al.}(2018)\citenamefont {Ade} \emph
  {et~al.}}]{Simons_Observatory}%
  \BibitemOpen
  \bibfield  {author} {\bibinfo {author} {\bibfnamefont {P.}~\bibnamefont
  {Ade}} \emph {et~al.} (\bibinfo {collaboration} {Simons Observatory}),\
  }\href@noop {} {\  (\bibinfo {year} {2018})},\ \Eprint
  {http://arxiv.org/abs/1808.07445} {arXiv:1808.07445 [astro-ph.CO]}
  \BibitemShut {NoStop}%
%%CITATION = ARXIV:1808.07445;%%
\bibitem [{\citenamefont {Enqvist}\ and\ \citenamefont
  {Sloth}(2002)}]{Enqvist:2001zp}%
  \BibitemOpen
  \bibfield  {author} {\bibinfo {author} {\bibfnamefont {K.}~\bibnamefont
  {Enqvist}}\ and\ \bibinfo {author} {\bibfnamefont {M.~S.}\ \bibnamefont
  {Sloth}},\ }\href {\doibase 10.1016/S0550-3213(02)00043-3} {\bibfield
  {journal} {\bibinfo  {journal} {Nucl. Phys.}\ }\textbf {\bibinfo {volume}
  {B626}},\ \bibinfo {pages} {395} (\bibinfo {year} {2002})},\ \Eprint
  {http://arxiv.org/abs/hep-ph/0109214} {arXiv:hep-ph/0109214 [hep-ph]}
  \BibitemShut {NoStop}%
%%CITATION = HEP-PH/0109214;%%
\bibitem [{\citenamefont {Lyth}\ and\ \citenamefont
  {Wands}(2002)}]{Lyth:2001nq}%
  \BibitemOpen
  \bibfield  {author} {\bibinfo {author} {\bibfnamefont {D.~H.}\ \bibnamefont
  {Lyth}}\ and\ \bibinfo {author} {\bibfnamefont {D.}~\bibnamefont {Wands}},\
  }\href {\doibase 10.1016/S0370-2693(01)01366-1} {\bibfield  {journal}
  {\bibinfo  {journal} {Phys. Lett.}\ }\textbf {\bibinfo {volume} {B524}},\
  \bibinfo {pages} {5} (\bibinfo {year} {2002})},\ \Eprint
  {http://arxiv.org/abs/hep-ph/0110002} {arXiv:hep-ph/0110002 [hep-ph]}
  \BibitemShut {NoStop}%
%%CITATION = HEP-PH/0110002;%%
\bibitem [{\citenamefont {Moroi}\ and\ \citenamefont
  {Takahashi}(2001)}]{Moroi:2001ct}%
  \BibitemOpen
  \bibfield  {author} {\bibinfo {author} {\bibfnamefont {T.}~\bibnamefont
  {Moroi}}\ and\ \bibinfo {author} {\bibfnamefont {T.}~\bibnamefont
  {Takahashi}},\ }\href {\doibase 10.1016/S0370-2693(02)02070-1,
  10.1016/S0370-2693(01)01295-3} {\bibfield  {journal} {\bibinfo  {journal}
  {Phys. Lett.}\ }\textbf {\bibinfo {volume} {B522}},\ \bibinfo {pages} {215}
  (\bibinfo {year} {2001})},\ \bibinfo {note} {[Erratum: Phys.
  Lett.B539,303(2002)]},\ \Eprint {http://arxiv.org/abs/hep-ph/0110096}
  {arXiv:hep-ph/0110096 [hep-ph]} \BibitemShut {NoStop}%
%%CITATION = HEP-PH/0110096;%%
\bibitem [{\citenamefont {Kawasaki}\ \emph {et~al.}(2000)\citenamefont
  {Kawasaki}, \citenamefont {Kohri},\ and\ \citenamefont
  {Sugiyama}}]{Kawasaki:2000en}%
  \BibitemOpen
  \bibfield  {author} {\bibinfo {author} {\bibfnamefont {M.}~\bibnamefont
  {Kawasaki}}, \bibinfo {author} {\bibfnamefont {K.}~\bibnamefont {Kohri}}, \
  and\ \bibinfo {author} {\bibfnamefont {N.}~\bibnamefont {Sugiyama}},\ }\href
  {\doibase 10.1103/PhysRevD.62.023506} {\bibfield  {journal} {\bibinfo
  {journal} {Phys. Rev.}\ }\textbf {\bibinfo {volume} {D62}},\ \bibinfo {pages}
  {023506} (\bibinfo {year} {2000})},\ \Eprint
  {http://arxiv.org/abs/astro-ph/0002127} {arXiv:astro-ph/0002127 [astro-ph]}
  \BibitemShut {NoStop}%
%%CITATION = ASTRO-PH/0002127;%%
\bibitem [{\citenamefont {Hannestad}(2004)}]{Hannestad:2004px}%
  \BibitemOpen
  \bibfield  {author} {\bibinfo {author} {\bibfnamefont {S.}~\bibnamefont
  {Hannestad}},\ }\href {\doibase 10.1103/PhysRevD.70.043506} {\bibfield
  {journal} {\bibinfo  {journal} {Phys. Rev.}\ }\textbf {\bibinfo {volume}
  {D70}},\ \bibinfo {pages} {043506} (\bibinfo {year} {2004})},\ \Eprint
  {http://arxiv.org/abs/astro-ph/0403291} {arXiv:astro-ph/0403291 [astro-ph]}
  \BibitemShut {NoStop}%
%%CITATION = ASTRO-PH/0403291;%%
\bibitem [{\citenamefont {Ichikawa}\ \emph {et~al.}(2005)\citenamefont
  {Ichikawa}, \citenamefont {Kawasaki},\ and\ \citenamefont
  {Takahashi}}]{Ichikawa:2005vw}%
  \BibitemOpen
  \bibfield  {author} {\bibinfo {author} {\bibfnamefont {K.}~\bibnamefont
  {Ichikawa}}, \bibinfo {author} {\bibfnamefont {M.}~\bibnamefont {Kawasaki}},
  \ and\ \bibinfo {author} {\bibfnamefont {F.}~\bibnamefont {Takahashi}},\
  }\href {\doibase 10.1103/PhysRevD.72.043522} {\bibfield  {journal} {\bibinfo
  {journal} {Phys. Rev.}\ }\textbf {\bibinfo {volume} {D72}},\ \bibinfo {pages}
  {043522} (\bibinfo {year} {2005})},\ \Eprint
  {http://arxiv.org/abs/astro-ph/0505395} {arXiv:astro-ph/0505395 [astro-ph]}
  \BibitemShut {NoStop}%
%%CITATION = ASTRO-PH/0505395;%%
\bibitem [{\citenamefont {De~Bernardis}\ \emph {et~al.}(2008)\citenamefont
  {De~Bernardis}, \citenamefont {Pagano},\ and\ \citenamefont
  {Melchiorri}}]{DeBernardis:2008zz}%
  \BibitemOpen
  \bibfield  {author} {\bibinfo {author} {\bibfnamefont {F.}~\bibnamefont
  {De~Bernardis}}, \bibinfo {author} {\bibfnamefont {L.}~\bibnamefont
  {Pagano}}, \ and\ \bibinfo {author} {\bibfnamefont {A.}~\bibnamefont
  {Melchiorri}},\ }\href {\doibase 10.1016/j.astropartphys.2008.09.005}
  {\bibfield  {journal} {\bibinfo  {journal} {Astropart. Phys.}\ }\textbf
  {\bibinfo {volume} {30}},\ \bibinfo {pages} {192} (\bibinfo {year}
  {2008})}\BibitemShut {NoStop}%
%%CITATION = APHYE,30,192;%%
\bibitem [{\citenamefont {Enckell}\ \emph {et~al.}(2019)\citenamefont
  {Enckell}, \citenamefont {Enqvist}, \citenamefont {Rasanen},\ and\
  \citenamefont {Wahlman}}]{Enckell:2018hmo}%
  \BibitemOpen
  \bibfield  {author} {\bibinfo {author} {\bibfnamefont {V.-M.}\ \bibnamefont
  {Enckell}}, \bibinfo {author} {\bibfnamefont {K.}~\bibnamefont {Enqvist}},
  \bibinfo {author} {\bibfnamefont {S.}~\bibnamefont {Rasanen}}, \ and\
  \bibinfo {author} {\bibfnamefont {L.-P.}\ \bibnamefont {Wahlman}},\ }\href
  {\doibase 10.1088/1475-7516/2019/02/022} {\bibfield  {journal} {\bibinfo
  {journal} {JCAP}\ }\textbf {\bibinfo {volume} {2019}},\ \bibinfo {pages}
  {022} (\bibinfo {year} {2019})},\ \Eprint {http://arxiv.org/abs/1810.05536}
  {arXiv:1810.05536 [gr-qc]} \BibitemShut {NoStop}%
%%CITATION = ARXIV:1810.05536;%%
\bibitem [{\citenamefont {Antoniadis}\ \emph {et~al.}(2018)\citenamefont
  {Antoniadis}, \citenamefont {Karam}, \citenamefont {Lykkas},\ and\
  \citenamefont {Tamvakis}}]{Antoniadis:2018ywb}%
  \BibitemOpen
  \bibfield  {author} {\bibinfo {author} {\bibfnamefont {I.}~\bibnamefont
  {Antoniadis}}, \bibinfo {author} {\bibfnamefont {A.}~\bibnamefont {Karam}},
  \bibinfo {author} {\bibfnamefont {A.}~\bibnamefont {Lykkas}}, \ and\ \bibinfo
  {author} {\bibfnamefont {K.}~\bibnamefont {Tamvakis}},\ }\href {\doibase
  10.1088/1475-7516/2018/11/028} {\bibfield  {journal} {\bibinfo  {journal}
  {JCAP}\ }\textbf {\bibinfo {volume} {1811}},\ \bibinfo {pages} {028}
  (\bibinfo {year} {2018})},\ \Eprint {http://arxiv.org/abs/1810.10418}
  {arXiv:1810.10418 [gr-qc]} \BibitemShut {NoStop}%
%%CITATION = ARXIV:1810.10418;%%
\bibitem [{\citenamefont {Antoniadis}\ \emph {et~al.}(2019)\citenamefont
  {Antoniadis}, \citenamefont {Karam}, \citenamefont {Lykkas}, \citenamefont
  {Pappas},\ and\ \citenamefont {Tamvakis}}]{Antoniadis:2018yfq}%
  \BibitemOpen
  \bibfield  {author} {\bibinfo {author} {\bibfnamefont {I.}~\bibnamefont
  {Antoniadis}}, \bibinfo {author} {\bibfnamefont {A.}~\bibnamefont {Karam}},
  \bibinfo {author} {\bibfnamefont {A.}~\bibnamefont {Lykkas}}, \bibinfo
  {author} {\bibfnamefont {T.}~\bibnamefont {Pappas}}, \ and\ \bibinfo {author}
  {\bibfnamefont {K.}~\bibnamefont {Tamvakis}},\ }\href {\doibase
  10.1088/1475-7516/2019/03/005} {\bibfield  {journal} {\bibinfo  {journal}
  {JCAP}\ }\textbf {\bibinfo {volume} {2019}},\ \bibinfo {pages} {005}
  (\bibinfo {year} {2019})},\ \Eprint {http://arxiv.org/abs/1812.00847}
  {arXiv:1812.00847 [gr-qc]} \BibitemShut {NoStop}%
%%CITATION = ARXIV:1812.00847;%%
\bibitem [{\citenamefont {Tenkanen}(2019{\natexlab{b}})}]{Tenkanen:2019jiq}%
  \BibitemOpen
  \bibfield  {author} {\bibinfo {author} {\bibfnamefont {T.}~\bibnamefont
  {Tenkanen}},\ }\href {\doibase 10.1103/PhysRevD.99.063528} {\bibfield
  {journal} {\bibinfo  {journal} {Phys. Rev.}\ }\textbf {\bibinfo {volume}
  {D99}},\ \bibinfo {pages} {063528} (\bibinfo {year} {2019}{\natexlab{b}})},\
  \Eprint {http://arxiv.org/abs/1901.01794} {arXiv:1901.01794 [astro-ph.CO]}
  \BibitemShut {NoStop}%
%%CITATION = ARXIV:1901.01794;%%
\bibitem [{\citenamefont {Birrell}\ and\ \citenamefont
  {Davies}(1984)}]{Birrell:1982ix}%
  \BibitemOpen
  \bibfield  {author} {\bibinfo {author} {\bibfnamefont {N.~D.}\ \bibnamefont
  {Birrell}}\ and\ \bibinfo {author} {\bibfnamefont {P.~C.~W.}\ \bibnamefont
  {Davies}},\ }\href {\doibase 10.1017/CBO9780511622632} {\emph {\bibinfo
  {title} {{Quantum Fields in Curved Space}}}},\ Cambridge Monographs on
  Mathematical Physics\ (\bibinfo  {publisher} {Cambridge Univ. Press},\
  \bibinfo {address} {Cambridge, UK},\ \bibinfo {year} {1984})\BibitemShut
  {NoStop}%
%%CITATION = INSPIRE-181166;%%
\bibitem [{\citenamefont {Tenkanen}(2020)}]{Tenkanen:2020dge}%
  \BibitemOpen
  \bibfield  {author} {\bibinfo {author} {\bibfnamefont {T.}~\bibnamefont
  {Tenkanen}},\ }\href@noop {} {\  (\bibinfo {year} {2020})},\ \Eprint
  {http://arxiv.org/abs/2001.10135} {arXiv:2001.10135 [astro-ph.CO]}
  \BibitemShut {NoStop}%
%%CITATION = ARXIV:2001.10135;%%
\bibitem [{\citenamefont {Sotiriou}\ and\ \citenamefont
  {Faraoni}(2010)}]{Sotiriou:2008rp}%
  \BibitemOpen
  \bibfield  {author} {\bibinfo {author} {\bibfnamefont {T.~P.}\ \bibnamefont
  {Sotiriou}}\ and\ \bibinfo {author} {\bibfnamefont {V.}~\bibnamefont
  {Faraoni}},\ }\href {\doibase 10.1103/RevModPhys.82.451} {\bibfield
  {journal} {\bibinfo  {journal} {Rev. Mod. Phys.}\ }\textbf {\bibinfo {volume}
  {82}},\ \bibinfo {pages} {451} (\bibinfo {year} {2010})},\ \Eprint
  {http://arxiv.org/abs/0805.1726} {arXiv:0805.1726 [gr-qc]} \BibitemShut
  {NoStop}%
%%CITATION = ARXIV:0805.1726;%%
\bibitem [{\citenamefont {Enckell}\ \emph {et~al.}(2020)\citenamefont
  {Enckell}, \citenamefont {Enqvist}, \citenamefont {Rasanen},\ and\
  \citenamefont {Wahlman}}]{Enckell:2018uic}%
  \BibitemOpen
  \bibfield  {author} {\bibinfo {author} {\bibfnamefont {V.-M.}\ \bibnamefont
  {Enckell}}, \bibinfo {author} {\bibfnamefont {K.}~\bibnamefont {Enqvist}},
  \bibinfo {author} {\bibfnamefont {S.}~\bibnamefont {Rasanen}}, \ and\
  \bibinfo {author} {\bibfnamefont {L.-P.}\ \bibnamefont {Wahlman}},\ }\href
  {\doibase 10.1088/1475-7516/2020/01/041} {\bibfield  {journal} {\bibinfo
  {journal} {JCAP}\ }\textbf {\bibinfo {volume} {2001}},\ \bibinfo {pages}
  {041} (\bibinfo {year} {2020})},\ \Eprint {http://arxiv.org/abs/1812.08754}
  {arXiv:1812.08754 [astro-ph.CO]} \BibitemShut {NoStop}%
%%CITATION = ARXIV:1812.08754;%%
\bibitem [{\citenamefont {Canko}\ \emph {et~al.}(2019)\citenamefont {Canko},
  \citenamefont {Gialamas},\ and\ \citenamefont {Kodaxis}}]{Canko:2019mud}%
  \BibitemOpen
  \bibfield  {author} {\bibinfo {author} {\bibfnamefont {D.~D.}\ \bibnamefont
  {Canko}}, \bibinfo {author} {\bibfnamefont {I.~D.}\ \bibnamefont {Gialamas}},
  \ and\ \bibinfo {author} {\bibfnamefont {G.~P.}\ \bibnamefont {Kodaxis}},\
  }\href@noop {} {\  (\bibinfo {year} {2019})},\ \Eprint
  {http://arxiv.org/abs/1901.06296} {arXiv:1901.06296 [hep-th]} \BibitemShut
  {NoStop}%
%%CITATION = ARXIV:1901.06296;%%
\bibitem [{\citenamefont {Rasanen}(2018)}]{Rasanen:2018ihz}%
  \BibitemOpen
  \bibfield  {author} {\bibinfo {author} {\bibfnamefont {S.}~\bibnamefont
  {Rasanen}},\ }\href {\doibase 10.21105/astro.1811.09514} {\bibfield
  {journal} {\bibinfo  {journal} {The Open Journal of Astrophysics}\ }
  (\bibinfo {year} {2018}),\ 10.21105/astro.1811.09514},\ \Eprint
  {http://arxiv.org/abs/1811.09514} {arXiv:1811.09514 [gr-qc]} \BibitemShut
  {NoStop}%
%%CITATION = ARXIV:1811.09514;%%
\bibitem [{\citenamefont {Järv}\ \emph {et~al.}(2018)\citenamefont {Järv},
  \citenamefont {Racioppi},\ and\ \citenamefont {Tenkanen}}]{Jarv:2017azx}%
  \BibitemOpen
  \bibfield  {author} {\bibinfo {author} {\bibfnamefont {L.}~\bibnamefont
  {Järv}}, \bibinfo {author} {\bibfnamefont {A.}~\bibnamefont {Racioppi}}, \
  and\ \bibinfo {author} {\bibfnamefont {T.}~\bibnamefont {Tenkanen}},\ }\href
  {\doibase 10.1103/PhysRevD.97.083513} {\bibfield  {journal} {\bibinfo
  {journal} {Phys. Rev.}\ }\textbf {\bibinfo {volume} {D97}},\ \bibinfo {pages}
  {083513} (\bibinfo {year} {2018})},\ \Eprint
  {http://arxiv.org/abs/1712.08471} {arXiv:1712.08471 [gr-qc]} \BibitemShut
  {NoStop}%
%%CITATION = ARXIV:1712.08471;%%
\bibitem [{\citenamefont {Azri}\ and\ \citenamefont
  {Demir}(2017)}]{Azri:2017uor}%
  \BibitemOpen
  \bibfield  {author} {\bibinfo {author} {\bibfnamefont {H.}~\bibnamefont
  {Azri}}\ and\ \bibinfo {author} {\bibfnamefont {D.}~\bibnamefont {Demir}},\
  }\href {\doibase 10.1103/PhysRevD.95.124007} {\bibfield  {journal} {\bibinfo
  {journal} {Phys. Rev.}\ }\textbf {\bibinfo {volume} {D95}},\ \bibinfo {pages}
  {124007} (\bibinfo {year} {2017})},\ \Eprint
  {http://arxiv.org/abs/1705.05822} {arXiv:1705.05822 [gr-qc]} \BibitemShut
  {NoStop}%
%%CITATION = ARXIV:1705.05822;%%
\bibitem [{\citenamefont {Azri}(2018)}]{Azri:2018gsz}%
  \BibitemOpen
  \bibfield  {author} {\bibinfo {author} {\bibfnamefont {H.}~\bibnamefont
  {Azri}},\ }\href {\doibase 10.1142/S0218271818300069} {\bibfield  {journal}
  {\bibinfo  {journal} {Int. J. Mod. Phys.}\ }\textbf {\bibinfo {volume}
  {D27}},\ \bibinfo {pages} {1830006} (\bibinfo {year} {2018})},\ \Eprint
  {http://arxiv.org/abs/1802.01247} {arXiv:1802.01247 [gr-qc]} \BibitemShut
  {NoStop}%
%%CITATION = ARXIV:1802.01247;%%
\bibitem [{\citenamefont {Shimada}\ \emph {et~al.}(2019)\citenamefont
  {Shimada}, \citenamefont {Aoki},\ and\ \citenamefont
  {Maeda}}]{Shimada:2018lnm}%
  \BibitemOpen
  \bibfield  {author} {\bibinfo {author} {\bibfnamefont {K.}~\bibnamefont
  {Shimada}}, \bibinfo {author} {\bibfnamefont {K.}~\bibnamefont {Aoki}}, \
  and\ \bibinfo {author} {\bibfnamefont {K.-i.}\ \bibnamefont {Maeda}},\ }\href
  {\doibase 10.1103/PhysRevD.99.104020} {\bibfield  {journal} {\bibinfo
  {journal} {Phys. Rev.}\ }\textbf {\bibinfo {volume} {D99}},\ \bibinfo {pages}
  {104020} (\bibinfo {year} {2019})},\ \Eprint
  {http://arxiv.org/abs/1812.03420} {arXiv:1812.03420 [gr-qc]} \BibitemShut
  {NoStop}%
%%CITATION = ARXIV:1812.03420;%%
\bibitem [{\citenamefont {Goldwirth}\ and\ \citenamefont
  {Piran}(1992)}]{GOLDWIRTH1992223}%
  \BibitemOpen
  \bibfield  {author} {\bibinfo {author} {\bibfnamefont {D.~S.}\ \bibnamefont
  {Goldwirth}}\ and\ \bibinfo {author} {\bibfnamefont {T.}~\bibnamefont
  {Piran}},\ }\href {\doibase https://doi.org/10.1016/0370-1573(92)90073-9}
  {\bibfield  {journal} {\bibinfo  {journal} {Physics Reports}\ }\textbf
  {\bibinfo {volume} {214}},\ \bibinfo {pages} {223 } (\bibinfo {year}
  {1992})}\BibitemShut {NoStop}%
\bibitem [{\citenamefont {Tenkanen}\ and\ \citenamefont
  {Tomberg}(2020)}]{Tenkanen:2020cvw}%
  \BibitemOpen
  \bibfield  {author} {\bibinfo {author} {\bibfnamefont {T.}~\bibnamefont
  {Tenkanen}}\ and\ \bibinfo {author} {\bibfnamefont {E.}~\bibnamefont
  {Tomberg}},\ }\href@noop {} {\  (\bibinfo {year} {2020})},\ \Eprint
  {http://arxiv.org/abs/2002.02420} {arXiv:2002.02420 [astro-ph.CO]}
  \BibitemShut {NoStop}%
%%CITATION = ARXIV:2002.02420;%%
\bibitem [{\citenamefont {Kaiser}\ and\ \citenamefont
  {Sfakianakis}(2014)}]{Kaiser:2013sna}%
  \BibitemOpen
  \bibfield  {author} {\bibinfo {author} {\bibfnamefont {D.~I.}\ \bibnamefont
  {Kaiser}}\ and\ \bibinfo {author} {\bibfnamefont {E.~I.}\ \bibnamefont
  {Sfakianakis}},\ }\href {\doibase 10.1103/PhysRevLett.112.011302} {\bibfield
  {journal} {\bibinfo  {journal} {Phys. Rev. Lett.}\ }\textbf {\bibinfo
  {volume} {112}},\ \bibinfo {pages} {011302} (\bibinfo {year} {2014})},\
  \Eprint {http://arxiv.org/abs/1304.0363} {arXiv:1304.0363 [astro-ph.CO]}
  \BibitemShut {NoStop}%
%%CITATION = ARXIV:1304.0363;%%
\bibitem [{\citenamefont {Carrilho}\ \emph {et~al.}(2018)\citenamefont
  {Carrilho}, \citenamefont {Mulryne}, \citenamefont {Ronayne},\ and\
  \citenamefont {Tenkanen}}]{Carrilho:2018ffi}%
  \BibitemOpen
  \bibfield  {author} {\bibinfo {author} {\bibfnamefont {P.}~\bibnamefont
  {Carrilho}}, \bibinfo {author} {\bibfnamefont {D.}~\bibnamefont {Mulryne}},
  \bibinfo {author} {\bibfnamefont {J.}~\bibnamefont {Ronayne}}, \ and\
  \bibinfo {author} {\bibfnamefont {T.}~\bibnamefont {Tenkanen}},\ }\href
  {\doibase 10.1088/1475-7516/2018/06/032} {\bibfield  {journal} {\bibinfo
  {journal} {JCAP}\ }\textbf {\bibinfo {volume} {1806}},\ \bibinfo {pages}
  {032} (\bibinfo {year} {2018})},\ \Eprint {http://arxiv.org/abs/1804.10489}
  {arXiv:1804.10489 [astro-ph.CO]} \BibitemShut {NoStop}%
%%CITATION = ARXIV:1804.10489;%%
\bibitem [{\citenamefont {Galante}\ \emph {et~al.}(2015)\citenamefont
  {Galante}, \citenamefont {Kallosh}, \citenamefont {Linde},\ and\
  \citenamefont {Roest}}]{Galante:2014ifa}%
  \BibitemOpen
  \bibfield  {author} {\bibinfo {author} {\bibfnamefont {M.}~\bibnamefont
  {Galante}}, \bibinfo {author} {\bibfnamefont {R.}~\bibnamefont {Kallosh}},
  \bibinfo {author} {\bibfnamefont {A.}~\bibnamefont {Linde}}, \ and\ \bibinfo
  {author} {\bibfnamefont {D.}~\bibnamefont {Roest}},\ }\href {\doibase
  10.1103/PhysRevLett.114.141302} {\bibfield  {journal} {\bibinfo  {journal}
  {Phys. Rev. Lett.}\ }\textbf {\bibinfo {volume} {114}},\ \bibinfo {pages}
  {141302} (\bibinfo {year} {2015})},\ \Eprint {http://arxiv.org/abs/1412.3797}
  {arXiv:1412.3797 [hep-th]} \BibitemShut {NoStop}%
%%CITATION = ARXIV:1412.3797;%%
\bibitem [{\citenamefont {Bauer}\ and\ \citenamefont
  {Demir}(2008)}]{Bauer:2008zj}%
  \BibitemOpen
  \bibfield  {author} {\bibinfo {author} {\bibfnamefont {F.}~\bibnamefont
  {Bauer}}\ and\ \bibinfo {author} {\bibfnamefont {D.~A.}\ \bibnamefont
  {Demir}},\ }\href {\doibase 10.1016/j.physletb.2008.06.014} {\bibfield
  {journal} {\bibinfo  {journal} {Phys. Lett.}\ }\textbf {\bibinfo {volume}
  {B665}},\ \bibinfo {pages} {222} (\bibinfo {year} {2008})},\ \Eprint
  {http://arxiv.org/abs/0803.2664} {arXiv:0803.2664 [hep-ph]} \BibitemShut
  {NoStop}%
%%CITATION = ARXIV:0803.2664;%%
\bibitem [{\citenamefont {Rasanen}\ and\ \citenamefont
  {Wahlman}(2017)}]{Rasanen:2017ivk}%
  \BibitemOpen
  \bibfield  {author} {\bibinfo {author} {\bibfnamefont {S.}~\bibnamefont
  {Rasanen}}\ and\ \bibinfo {author} {\bibfnamefont {P.}~\bibnamefont
  {Wahlman}},\ }\href {\doibase 10.1088/1475-7516/2017/11/047} {\bibfield
  {journal} {\bibinfo  {journal} {JCAP}\ }\textbf {\bibinfo {volume} {1711}},\
  \bibinfo {pages} {047} (\bibinfo {year} {2017})},\ \Eprint
  {http://arxiv.org/abs/1709.07853} {arXiv:1709.07853 [astro-ph.CO]}
  \BibitemShut {NoStop}%
%%CITATION = ARXIV:1709.07853;%%
\bibitem [{\citenamefont {Takahashi}\ and\ \citenamefont
  {Tenkanen}(2019)}]{Takahashi:2018brt}%
  \BibitemOpen
  \bibfield  {author} {\bibinfo {author} {\bibfnamefont {T.}~\bibnamefont
  {Takahashi}}\ and\ \bibinfo {author} {\bibfnamefont {T.}~\bibnamefont
  {Tenkanen}},\ }\href {\doibase 10.1088/1475-7516/2019/04/035} {\bibfield
  {journal} {\bibinfo  {journal} {JCAP}\ }\textbf {\bibinfo {volume} {1904}},\
  \bibinfo {pages} {035} (\bibinfo {year} {2019})},\ \Eprint
  {http://arxiv.org/abs/1812.08492} {arXiv:1812.08492 [astro-ph.CO]}
  \BibitemShut {NoStop}%
%%CITATION = ARXIV:1812.08492;%%
\bibitem [{\citenamefont {Cosme}\ \emph
  {et~al.}(2018{\natexlab{b}})\citenamefont {Cosme}, \citenamefont {Rosa},\
  and\ \citenamefont {Bertolami}}]{Cosme:2018nly}%
  \BibitemOpen
  \bibfield  {author} {\bibinfo {author} {\bibfnamefont {C.}~\bibnamefont
  {Cosme}}, \bibinfo {author} {\bibfnamefont {J.~G.}\ \bibnamefont {Rosa}}, \
  and\ \bibinfo {author} {\bibfnamefont {O.}~\bibnamefont {Bertolami}},\ }\href
  {\doibase 10.1007/JHEP05(2018)129} {\bibfield  {journal} {\bibinfo  {journal}
  {JHEP}\ }\textbf {\bibinfo {volume} {05}},\ \bibinfo {pages} {129} (\bibinfo
  {year} {2018}{\natexlab{b}})},\ \Eprint {http://arxiv.org/abs/1802.09434}
  {arXiv:1802.09434 [hep-ph]} \BibitemShut {NoStop}%
%%CITATION = ARXIV:1802.09434;%%
\bibitem [{\citenamefont {Enqvist}\ \emph {et~al.}(2014)\citenamefont
  {Enqvist}, \citenamefont {Nurmi}, \citenamefont {Tenkanen},\ and\
  \citenamefont {Tuominen}}]{Enqvist:2014zqa}%
  \BibitemOpen
  \bibfield  {author} {\bibinfo {author} {\bibfnamefont {K.}~\bibnamefont
  {Enqvist}}, \bibinfo {author} {\bibfnamefont {S.}~\bibnamefont {Nurmi}},
  \bibinfo {author} {\bibfnamefont {T.}~\bibnamefont {Tenkanen}}, \ and\
  \bibinfo {author} {\bibfnamefont {K.}~\bibnamefont {Tuominen}},\ }\href
  {\doibase 10.1088/1475-7516/2014/08/035} {\bibfield  {journal} {\bibinfo
  {journal} {JCAP}\ }\textbf {\bibinfo {volume} {1408}},\ \bibinfo {pages}
  {035} (\bibinfo {year} {2014})},\ \Eprint {http://arxiv.org/abs/1407.0659}
  {arXiv:1407.0659 [astro-ph.CO]} \BibitemShut {NoStop}%
%%CITATION = ARXIV:1407.0659;%%
\bibitem [{\citenamefont {Kainulainen}\ \emph {et~al.}(2016)\citenamefont
  {Kainulainen}, \citenamefont {Nurmi}, \citenamefont {Tenkanen}, \citenamefont
  {Tuominen},\ and\ \citenamefont {Vaskonen}}]{Kainulainen:2016vzv}%
  \BibitemOpen
  \bibfield  {author} {\bibinfo {author} {\bibfnamefont {K.}~\bibnamefont
  {Kainulainen}}, \bibinfo {author} {\bibfnamefont {S.}~\bibnamefont {Nurmi}},
  \bibinfo {author} {\bibfnamefont {T.}~\bibnamefont {Tenkanen}}, \bibinfo
  {author} {\bibfnamefont {K.}~\bibnamefont {Tuominen}}, \ and\ \bibinfo
  {author} {\bibfnamefont {V.}~\bibnamefont {Vaskonen}},\ }\href {\doibase
  10.1088/1475-7516/2016/06/022} {\bibfield  {journal} {\bibinfo  {journal}
  {JCAP}\ }\textbf {\bibinfo {volume} {1606}},\ \bibinfo {pages} {022}
  (\bibinfo {year} {2016})},\ \Eprint {http://arxiv.org/abs/1601.07733}
  {arXiv:1601.07733 [astro-ph.CO]} \BibitemShut {NoStop}%
%%CITATION = ARXIV:1601.07733;%%
\bibitem [{\citenamefont {Heikinheimo}\ \emph {et~al.}(2016)\citenamefont
  {Heikinheimo}, \citenamefont {Tenkanen}, \citenamefont {Tuominen},\ and\
  \citenamefont {Vaskonen}}]{Heikinheimo:2016yds}%
  \BibitemOpen
  \bibfield  {author} {\bibinfo {author} {\bibfnamefont {M.}~\bibnamefont
  {Heikinheimo}}, \bibinfo {author} {\bibfnamefont {T.}~\bibnamefont
  {Tenkanen}}, \bibinfo {author} {\bibfnamefont {K.}~\bibnamefont {Tuominen}},
  \ and\ \bibinfo {author} {\bibfnamefont {V.}~\bibnamefont {Vaskonen}},\
  }\href {\doibase 10.1103/PhysRevD.96.109902, 10.1103/PhysRevD.94.063506}
  {\bibfield  {journal} {\bibinfo  {journal} {Phys. Rev.}\ }\textbf {\bibinfo
  {volume} {D94}},\ \bibinfo {pages} {063506} (\bibinfo {year} {2016})},\
  \bibinfo {note} {[Erratum: Phys. Rev.D96,no.10,109902(2017)]},\ \Eprint
  {http://arxiv.org/abs/1604.02401} {arXiv:1604.02401 [astro-ph.CO]}
  \BibitemShut {NoStop}%
%%CITATION = ARXIV:1604.02401;%%
\bibitem [{\citenamefont {Enqvist}\ \emph {et~al.}(2018)\citenamefont
  {Enqvist}, \citenamefont {Hardwick}, \citenamefont {Tenkanen}, \citenamefont
  {Vennin},\ and\ \citenamefont {Wands}}]{Enqvist:2017kzh}%
  \BibitemOpen
  \bibfield  {author} {\bibinfo {author} {\bibfnamefont {K.}~\bibnamefont
  {Enqvist}}, \bibinfo {author} {\bibfnamefont {R.~J.}\ \bibnamefont
  {Hardwick}}, \bibinfo {author} {\bibfnamefont {T.}~\bibnamefont {Tenkanen}},
  \bibinfo {author} {\bibfnamefont {V.}~\bibnamefont {Vennin}}, \ and\ \bibinfo
  {author} {\bibfnamefont {D.}~\bibnamefont {Wands}},\ }\href {\doibase
  10.1088/1475-7516/2018/02/006} {\bibfield  {journal} {\bibinfo  {journal}
  {JCAP}\ }\textbf {\bibinfo {volume} {1802}},\ \bibinfo {pages} {006}
  (\bibinfo {year} {2018})},\ \Eprint {http://arxiv.org/abs/1711.07344}
  {arXiv:1711.07344 [astro-ph.CO]} \BibitemShut {NoStop}%
%%CITATION = ARXIV:1711.07344;%%
\bibitem [{\citenamefont {Takahashi}\ \emph {et~al.}(2018)\citenamefont
  {Takahashi}, \citenamefont {Yin},\ and\ \citenamefont {Guth}}]{Guth:2018hsa}%
  \BibitemOpen
  \bibfield  {author} {\bibinfo {author} {\bibfnamefont {F.}~\bibnamefont
  {Takahashi}}, \bibinfo {author} {\bibfnamefont {W.}~\bibnamefont {Yin}}, \
  and\ \bibinfo {author} {\bibfnamefont {A.~H.}\ \bibnamefont {Guth}},\ }\href
  {\doibase 10.1103/PhysRevD.98.015042} {\bibfield  {journal} {\bibinfo
  {journal} {Phys. Rev.}\ }\textbf {\bibinfo {volume} {D98}},\ \bibinfo {pages}
  {015042} (\bibinfo {year} {2018})},\ \Eprint
  {http://arxiv.org/abs/1805.08763} {arXiv:1805.08763 [hep-ph]} \BibitemShut
  {NoStop}%
%%CITATION = ARXIV:1805.08763;%%
\bibitem [{\citenamefont {Graham}\ and\ \citenamefont
  {Scherlis}(2018)}]{Graham:2018jyp}%
  \BibitemOpen
  \bibfield  {author} {\bibinfo {author} {\bibfnamefont {P.~W.}\ \bibnamefont
  {Graham}}\ and\ \bibinfo {author} {\bibfnamefont {A.}~\bibnamefont
  {Scherlis}},\ }\href {\doibase 10.1103/PhysRevD.98.035017} {\bibfield
  {journal} {\bibinfo  {journal} {Phys. Rev.}\ }\textbf {\bibinfo {volume}
  {D98}},\ \bibinfo {pages} {035017} (\bibinfo {year} {2018})},\ \Eprint
  {http://arxiv.org/abs/1805.07362} {arXiv:1805.07362 [hep-ph]} \BibitemShut
  {NoStop}%
%%CITATION = ARXIV:1805.07362;%%
\bibitem [{\citenamefont {Bunch}\ and\ \citenamefont
  {Davies}(1978)}]{Bunch:1978yq}%
  \BibitemOpen
  \bibfield  {author} {\bibinfo {author} {\bibfnamefont {T.~S.}\ \bibnamefont
  {Bunch}}\ and\ \bibinfo {author} {\bibfnamefont {P.~C.~W.}\ \bibnamefont
  {Davies}},\ }\href {\doibase 10.1098/rspa.1978.0060} {\bibfield  {journal}
  {\bibinfo  {journal} {Proc. Roy. Soc. Lond.}\ }\textbf {\bibinfo {volume}
  {A360}},\ \bibinfo {pages} {117} (\bibinfo {year} {1978})}\BibitemShut
  {NoStop}%
%%CITATION = PRSLA,A360,117;%%
\bibitem [{\citenamefont {Vilenkin}\ and\ \citenamefont
  {Ford}(1982)}]{Vilenkin:1982wt}%
  \BibitemOpen
  \bibfield  {author} {\bibinfo {author} {\bibfnamefont {A.}~\bibnamefont
  {Vilenkin}}\ and\ \bibinfo {author} {\bibfnamefont {L.~H.}\ \bibnamefont
  {Ford}},\ }\href {\doibase 10.1103/PhysRevD.26.1231} {\bibfield  {journal}
  {\bibinfo  {journal} {Phys. Rev.}\ }\textbf {\bibinfo {volume} {D26}},\
  \bibinfo {pages} {1231} (\bibinfo {year} {1982})}\BibitemShut {NoStop}%
%%CITATION = PHRVA,D26,1231;%%
\bibitem [{\citenamefont {Starobinsky}\ and\ \citenamefont
  {Yokoyama}(1994)}]{Starobinsky:1994bd}%
  \BibitemOpen
  \bibfield  {author} {\bibinfo {author} {\bibfnamefont {A.~A.}\ \bibnamefont
  {Starobinsky}}\ and\ \bibinfo {author} {\bibfnamefont {J.}~\bibnamefont
  {Yokoyama}},\ }\href {\doibase 10.1103/PhysRevD.50.6357} {\bibfield
  {journal} {\bibinfo  {journal} {Phys. Rev.}\ }\textbf {\bibinfo {volume}
  {D50}},\ \bibinfo {pages} {6357} (\bibinfo {year} {1994})},\ \Eprint
  {http://arxiv.org/abs/astro-ph/9407016} {arXiv:astro-ph/9407016 [astro-ph]}
  \BibitemShut {NoStop}%
%%CITATION = ASTRO-PH/9407016;%%
\bibitem [{\citenamefont {Markkanen}\ \emph {et~al.}(2019)\citenamefont
  {Markkanen}, \citenamefont {Rajantie}, \citenamefont {Stopyra},\ and\
  \citenamefont {Tenkanen}}]{Markkanen:2019kpv}%
  \BibitemOpen
  \bibfield  {author} {\bibinfo {author} {\bibfnamefont {T.}~\bibnamefont
  {Markkanen}}, \bibinfo {author} {\bibfnamefont {A.}~\bibnamefont {Rajantie}},
  \bibinfo {author} {\bibfnamefont {S.}~\bibnamefont {Stopyra}}, \ and\
  \bibinfo {author} {\bibfnamefont {T.}~\bibnamefont {Tenkanen}},\ }\href
  {\doibase 10.1088/1475-7516/2019/08/001} {\bibfield  {journal} {\bibinfo
  {journal} {JCAP}\ }\textbf {\bibinfo {volume} {1908}},\ \bibinfo {pages}
  {001} (\bibinfo {year} {2019})},\ \Eprint {http://arxiv.org/abs/1904.11917}
  {arXiv:1904.11917 [gr-qc]} \BibitemShut {NoStop}%
%%CITATION = ARXIV:1904.11917;%%
\bibitem [{\citenamefont {Enqvist}\ \emph {et~al.}(2012)\citenamefont
  {Enqvist}, \citenamefont {Lerner}, \citenamefont {Taanila},\ and\
  \citenamefont {Tranberg}}]{Enqvist:2012xn}%
  \BibitemOpen
  \bibfield  {author} {\bibinfo {author} {\bibfnamefont {K.}~\bibnamefont
  {Enqvist}}, \bibinfo {author} {\bibfnamefont {R.~N.}\ \bibnamefont {Lerner}},
  \bibinfo {author} {\bibfnamefont {O.}~\bibnamefont {Taanila}}, \ and\
  \bibinfo {author} {\bibfnamefont {A.}~\bibnamefont {Tranberg}},\ }\href
  {\doibase 10.1088/1475-7516/2012/10/052} {\bibfield  {journal} {\bibinfo
  {journal} {JCAP}\ }\textbf {\bibinfo {volume} {1210}},\ \bibinfo {pages}
  {052} (\bibinfo {year} {2012})},\ \Eprint {http://arxiv.org/abs/1205.5446}
  {arXiv:1205.5446 [astro-ph.CO]} \BibitemShut {NoStop}%
%%CITATION = ARXIV:1205.5446;%%
\end{thebibliography}%

%%%%%%%%%%%%%%%%%%%%%%%%%%%%%%%%%%%%%%%%%%%%%%%%%%%%%%%%%%%%%%%%%%%%%%%%%%%%%%%%%%%%%%%%%%

\end{document}